\documentclass[10pt,letterpaper]{article}
\usepackage[top=0.85in,left=2.75in,footskip=0.75in]{geometry}

\usepackage{graphicx}
\usepackage{float}
\usepackage{array}
\usepackage{geometry}
\usepackage{multirow}
\usepackage{makecell}

\usepackage{adjustbox}

\usepackage{amsmath,amssymb}
\usepackage{amsfonts} 
\usepackage{lmodern} 
\usepackage[T1]{fontenc}
\usepackage{textcomp}

\usepackage{changepage}

\usepackage[utf8x]{inputenc}

\usepackage{textcomp,marvosym}

\usepackage{cite}

\usepackage{nameref,hyperref}

\usepackage[right]{lineno}

\usepackage{microtype}
\DisableLigatures[f]{encoding = *, family = * }

\usepackage[table]{xcolor}

\usepackage{array}

\newcolumntype{+}{!{\vrule width 2pt}}

\newlength\savedwidth

\raggedright
\usepackage[aboveskip=1pt,labelfont=bf,labelsep=period,justification=raggedright,singlelinecheck=off]{caption}

\makeatletter
\renewcommand{\@biblabel}[1]{\quad#1.}
\makeatother

\usepackage{lastpage,fancyhdr,graphicx}
\usepackage{epstopdf}
\fancyheadoffset[L]{2.25in}
\fancyfootoffset[L]{2.25in}
\definecolor{lightergray}{rgb}{0.9, 0.9, 0.9} 
\definecolor{lightred}{rgb}{1, 0.4, 0.4} 
\definecolor{lightorange}{rgb}{1, 0.7, 0.5} 

\begin{document}
\vspace*{0.2in}

\begin{flushleft}
{\Large
\textbf\newline{The impact of labeling automotive AI as ``trustworthy'' or ``reliable'' on user evaluation and technology acceptance} 
}
\newline
\\
John Dorsch\textsuperscript{1*},
Ophelia Deroy\textsuperscript{1**}
\\
\bigskip
\textbf{1} Faculty of Philosophy, Philosophy of Science and the Study of Religion, Ludwig-Maximilians-Universität München, Munich, Germany
\\
\bigskip

* \href{mailto:johndorsch@gmail.com}{johndorsch@gmail.com} \\
** \href{mailto:ophelia.deroy@example.com}{ophelia.deroy@lmu.de}

\end{flushleft}


\section*{Abstract}
This study explores whether labeling AI as either ``trustworthy'' or ``reliable'' influences user perceptions and acceptance of automotive AI technologies. Utilizing a one-way between-subjects design, the research presented online participants (N=478) with a text presenting guidelines for either trustworthy or reliable AI, before asking them to evaluate 3 vignette scenarios and fill in a modified version of the Technology Acceptance Model which covers different variables, such as perceived ease of use, human-like trust, and overall attitude. While labeling AI as ``trustworthy'' did not significantly influence people's judgements on specific scenarios, it increased perceived ease of use and human-like trust, namely benevolence, suggesting a facilitating influence on usability and an anthropomorphic effect on user perceptions. The study provides insights into how specific labels affect adopting certain attitudes toward AI technology. 


\section*{Introduction}

The wide use of the "trustworthy AI" denomination by manufacturers and legislators sparks debates. On the one hand, several philosophers have argued that it is irrational to place trust in a machine \cite{bib1, bib2, bib3, bib20}. Instead, trust should be placed not in the machines themselves, but in the humans responsible for their development and deployment \cite{bib4}. Others, like Coeckelbergh \cite{bib5} for instance, argues in favor of trusting machines, emphasizing that as social animals, we already live in a world structured by trusting relationships. Accordingly, trust should be considered a pre-reflective condition of human experience and unproblematically extended to machines. 

Those who argue against trusting machines contend that trust can only be rationally employed if the trustee is sensitive to normative constraints of reasons \cite{bib3}, has the trustee’s goodwill in mind when acting \cite{bib6}, holds the trustee’s interests as their own \cite{bib7}, or is emotionally sensitive to moral values \cite{bib8}. Since today's machines are nowehere near  any of these conditions, many AI ethicists advocate for employing the notion of reliability rather than trust when dealing with machines. Reliability, decoupled from trust, is understood as entailing solely performance-based or outcome-based standards for evaluation \cite{bib2} \cite{bib3}.

But do these distinct conceptual frameworks for conceiving of one’s relationship to machines- the framework of trust and the framework of reliability - influence the attitudes of users towards this technology? 

In this study, we sought to test whether the deployment of either framework in the field of automotive AI technology would predict changes in naive users attitudes. Specifically, we investigated whether either framework would predict changes in attitudes toward automotive AI regarding both specific scenarios and general dimensions of technological acceptance. In specific scenarios where AI technologies lead to a poor outcome, humans tend to hold machines responsible, that is, blame and praise them \cite{bib9, bib10, bib11}. As these are more anthropocentric actions with social ramifications, the question arises whether humans will tend to hold AI less responsible when employing a reliability-based framework compared to a trust-based one. 

When it comes to technology acceptance, empirical investigation often relies on the Technology Acceptance Model (TAM). The TAM framework helps in understanding how users come to accept and use a technology, focusing on distinct factors such as perceived ease of use, perceived usefulness, and attitude towards the technology. In our study, we employed a streamlined TAM based on the constructs identified by Davis \cite{bib12} and further refined by Venkatesh and Davis \cite{bib13}, as well as insights from Choung et al. \cite{bib14}, to assess whether either the trust or reliability framework would predict changes in these variables. This approach allowed us to measure whether these frameworks would influence overall technology acceptance or specific aspects. 

Previous studies investigating user attitudes toward automotive AI have had as their focus anthropomorphized autonomous vehicles \cite{bib15, bib16, bib17, bib18}, showing that enhanced anthropomorphization can increase perceived trust in this technology (cf. \cite{bib21}). Anthropomorphizing likely skews participant responses toward trustworthiness, since the more anthropomorphized the autonomous car is, the more inclined the participant will be to treat the technology like a human-like agent. For this reason, this study seeks to understand attitudes toward AI assistance in driving, wherein it is arguably more appropriate to conceive of AI as a tool. 

Finally, understanding the effects of labels on technology acceptance is worthwhile, as these can offer insight into how to mitigate algorithm aversion or exploitation—phenomena where users reject AI technology despite its potential to enhance performance or collective benefits \cite{bib19, bib21, bib22}. 

\subsubsection*{Hypotheses}

In this study, we investigated the impact of labeling AI as ``trustworthy'' versus ``reliable'' on various dependent variables related to technology attitudes. Our hypotheses (\textbf{H1}, \textbf{H2}, \textbf{H3}, \textbf{H4}, \textbf{H5}) examine whether the label ``trustworthy AI'' predicts changes in AI blameworthiness, AI accountability, confidence while using, and confidence while learning, and the overall TAM score. Additionally, we explored whether the labels predict changes in TAM categories, including perceived ease of use, perceived usefulness, intention to use,  ability-based and human-like trust in automotive AI (\textbf{H5.1 - H5.8}) (see Table \ref{table:hypotheses}). The study was pre-registered to enhance transparency and credibility. The hypotheses remain consistent with those outlined in the pre-registration, though the numbering has been reordered and simplified, the null hypotheses are reformulated as `no predicted effect'. The pre-registration document can be accessed at the Open Science Framework (OSF) via \href{https://osf.io/6un4h/}{this link}.

\section*{Materials and methods}
\subsection*{Participants}

Initially, 617 participants were recruited for the study through online platforms Cloud Research and Amazon Mechanical Turk. Participants were paid approximately \$12 an hour for completing the study, whether their data was ultimately analyzed or not. After conducting a language proficiency check, 66 participants were excluded, leaving 551 (see \nameref{lang}). An additional 58 participants were excluded due to failing the attention check, resulting in 493 participants (see \nameref{att}). A further 15 participants were excluded based on age criteria, leaving a final sample of 478 participants whose data were analyzed. The majority of participants were either male (246) or female (242). There were also 3 participants who identified as non-binary, 1 participant who preferred not to say, and 1 who identified as `other' (for more, see \nameref{demo}).

As outlined in the pre-registered plan, the sample size for the study was determined through a power analysis, ensuring sufficient statistical power to detect meaningful effects in the primary dependent variables. To account for potential exclusions due to language and attention check failures, an additional 10\% was added to the calculated sample size, leading to a planned total of 440 participants. For the variables corresponding to the vignettes (Hypotheses \textbf{H1} to \textbf{H4}), the power analysis was conducted using analytic methods based on effect sizes calculated from Cohen’s d, derived from the pilot study data on 5-point Likert scale variables. The analysis indicated that a sample size of 400 participants would yield sufficient power \( (0.80) \) to detect the observed effect sizes at a significance level of \( \alpha = 0.05 \).

For the TAM questionnaire (\textbf{H5} and \textbf{H5.1 - H5.8}), the power analysis was based on pilot study data. The effect sizes observed from the pilot study for the 5-point Likert scale variables were \( -0.193 \), \( -0.257 \), \( -0.507 \), and \( -0.431 \). These were used alongside threshold estimates and their standard errors, which were \( 4.8725 \) \((\text{SE} = 0.5113)\), \( 3.5229 \) \((\text{SE} = 0.4330)\), \( 1.2539 \) \((\text{SE} = 0.3788)\), and \( -1.7388 \) \((\text{SE} = 0.3885)\), respectively. The standard deviations of participants and items were \( 1.4363 \) and \( 0.3499 \), respectively. With these parameters, a sample size of 400 participants was determined to provide a power of \( 0.80 \) at a significance level of \( \alpha = 0.05 \). This was confirmed through simulation involving 100 iterations, each with 400 participants, ensuring robustness in the power estimate.

During data collection, an initial passing sample of 390 participants revealed a gender imbalance, with approximately 65\% male and 35\% female participants (245 male participants and 145 female participants). To achieve a balanced gender demographic as outlined in pre-registration, additional data were collected exclusively from women, resulting in a more balanced sample. Our preliminary analysis indicated that gender, specifically being female, significantly influenced attitudes toward automotive AI, with women reporting more negative attitudes. This finding underscored the importance of ensuring a balanced sample to accurately capture the diversity of perspectives (for more detailed information, see \nameref{gender}).

We categorized participants into three age groups for analysis: 18-30, 30-45, and 45-65. Due to the nature of these categories, there is potential ambiguity at the boundaries, specifically for participants aged exactly 30 and 45. In our dataset, we could not precisely allocate these boundary ages to a specific group. Consequently, there may be classification ambiguity at these boundaries, which could influence the interpretation of age-related findings (for more, see \nameref{age}).

\subsection*{Experimental Design}
A one-way between-subjects design was used. Participants were randomly assigned to one of two groups: the ``trustworthy AI'' group or the ``reliable AI'' group. The assignment was random to ensure that each participant had an equal chance of being placed in either group. Ultimately, data were analyzing from 241 participants assigned to the ``trustworthy AI'' group and 237 participants assigned to the ``reliable AI'' group, ensuring a balanced sample.  The sample consisted of 493 participants, with a balanced distribution in terms of gender and age groups. As specified in the pre-registered plan, 15 participants were excluded from analysis due to being 65 years old or older, leaving a final sample of 478 participants whose data were analyzed. The majority of participants were either male (246) or female (242). There were also 3 participants who identified as non-binary, 1 participant who preferred not to say, and 1 who identified as `other'. The age distribution was as follows: 68 participants were 18-30 years old, 240 were 30-45 years old, 167 were 45-64 years old, 15 were 65 years or older, and 3 preferred not to say. A vast majority of participants held a driver's license (476), while 14 participants did not, and 3 preferred not to disclose their status.

After group assignment, participants progressed through 4 phases.

\subsubsection*{Phase 1: Demographic Phase}
This section involved the collection of demographic data (age, gender, education, experience with automotive AI, driver’s license status, AI expertise), and a language check to ensure fluency in English. Detailed demographic data are described in \nameref{demo}, and language check criteria are provided in \nameref{supplementary_material}, specifically in \nameref{lang}.

\subsubsection*{Phase 2: Induction Phase}
Participants were presented with group-specific definitions of ``trustworthy AI'' or ``reliable AI'' (see \nameref{defs}) attributed to AI experts, as well as an attention check to ensure engagement and comprehension of the respective definition (see \nameref{att}).

\subsubsection*{Phase 3: Vignette Phase}
Participants read three separate vignettes, one for each automotive AI-assisted task: planning assistance, parking assistance, and steering assistance. The vignettes were presented in random order and, across the two groups, the vignettes were identical except for the key terms ``trustworthy'' or ``reliable'' (and other related semantic derivatives) respective of group assignment. Participants answered the same four questions for each of the three vignettes (AI accountability, AI blameworthiness, confidence in driving, and confidence in learning) using a 5-point Likert scale, which were also presented in random order. The full text of the vignettes and the questions asked are available in \nameref{supplementary_material}, specifically in \nameref{vign}.

\subsubsection*{Phase 4: Questionnaire Phase}
Participants answered the eight TAM questions covering the distinct constructs perceived ease of use, perceived usefulness, behavioral intention, ability trust, human-like trust (benevolence and integrity), general trust, and attitude. A 5-point Likert scale was used to measure responses. The questions were presented in random order to prevent biasing effects. The complete TAM questionnaire is provided in \nameref{supplementary_material}, specifically in \nameref{TAM_Qs}.
\newline
 
The selection of items for TAM is informed by Choung et al. \cite{bib14}, which identified the importance of trust (general, human-like, and ability-based trust) in AI and its impact on technology acceptance. Their comprehensive study provides a validated framework for measuring key constructs related to technology acceptance. Each TAM variable category was measured with one question to streamline the data collection process, minimize participant fatigue, and cohere with automotive AI. 

The selected questions are the most appropriate for automotive AI, and the wording of each question has been modified to reflect this technology. The original study asked about smart AI assistants and, as a result, several of the survey items do not apply. The human-like trust variable is assessed using two questions as in the original study in order to ensure a comprehensive and reliable measurement. This is necessary due to the complexity and multidimensional nature of human-like trust, which includes two distinct elements, benevolence and integrity. 

\subsection*{Data Collection and Handling}
Data was collected using the Qualtrics online survey platform. Procedures for data handling and storage included secure, encrypted storage and removal of personal identifiers. Data monitoring and quality assurance were ensured through exclusion of incomplete responses, failed language checks, and failed attention checks. See \nameref{supplementary_material} for more, specifically \nameref{inc}.

\subsection*{Data Analysis}

For the analysis of the vignettes, independent samples t-tests were conducted to compare means by condition for hypotheses related to the vignette assessments, i.e. \textbf{H1} to \textbf{H4}, which include AI accountability, AI blameworthiness, confidence in using automotive AI, and confidence in learning how to drive with automotive AI (in the context of obtaining one's first driver's license). To further investigate the effects of condition on outcomes related to the vignettes, Bayesian ordinal regression models were fitted for each question. These models accounted for participant and vignette as random effects, providing posterior estimates of the condition effect for each hypothesis. 

For the total TAM score, addressing hypothesis \textbf{H5}, we employed a Cumulative Link Mixed Model (CLMM) to account for the ordinal nature of the response variable and to model participant and TAM items as random effects. Cumulative Link Models (CLMs) were used to analyze ordinal response variables related to the individual TAM questions, specifically testing hypotheses \textbf{H5.1} through \textbf{H5.8}.

Analysis deviated slightly from the pre-registered plan. First, the residuals of the t-test did not adhere to a normal distribution, as indicated by significant deviations in the Shapiro-Wilk test results. Since the t-test assumes normally distributed residuals, its application was questioned for this data. We still performed the t-test and report the results below. To address the non-normality of residuals, we also conducted a Wilcoxon test as a non-parametric alternative; however, this test did not reveal any significant results either. Detailed results of the Wilcoxon test are reported in \nameref{supplementary_material}, respectively in \nameref{T-Test_Sup}. Finally, we opted not to use Cumulative Link Mixed Models for \textbf{H5.1} through \textbf{H5.8} as initially proposed. Since we modelled each question separately, we did not need to account for random effects of items or participants, so a simpler ordinal regression model (CLM) was sufficient for analysis.

\section*{Results}


\begin{table}[ht]
\centering
\caption{\textbf{Hypotheses and Results Overview}}
\caption*{\footnotesize Each hypothesis pertains to the label  
``trustworthy AI'' and its potential effect on the variable (increase or decrease). Red lines indicate an observed effect that was different from the predicted effect. Orange indicates that the null hypothesis could not be rejected. No predicted effects were observed.}
\vspace{1em}
\label{table:hypotheses}
\begin{tabular}{|>{\columncolor{lightergray}}c|c|c|c|c|}
\hline
\rowcolor{lightgray} \textbf{} & 
\textbf{Variable} & 
\makecell{\textbf{Predicted}\\ \textbf{Difference}} & 
\makecell{\textbf{No}\\ \textbf{Prediction}} & 
\makecell{\textbf{Observed}\\ \textbf{Difference}} \\ 
\hline
\rowcolor{lightorange} \textbf{H1} & \textbf{AI Blameworthiness} & \textbf{decrease} &  & \textbf{none} \\ 
\hline
\rowcolor{lightorange} \textbf{H2} & \textbf{AI Accountability} & \textbf{decrease} &  & \textbf{none} \\ 
\hline
\textbf{H3} & Confidence in Using & & \checkmark & none \\ 
\hline
\textbf{H4} & Confidence in Learning & & \checkmark & none \\ 
\hline
\rowcolor{lightorange} \textbf{H5} & \textbf{Total TAM Score} & \textbf{decrease} & & \textbf{none} \\ 
\hline
\rowcolor{lightred} \textbf{H5.1} & \textbf{Ease of Use} & & \textbf{\checkmark} & \textbf{increase} \\ 
\hline
\rowcolor{lightorange} \textbf{H5.2} & \textbf{Usefulness} & \textbf{decrease} & & \textbf{none} \\ 
\hline
\textbf{H5.3} & Intention to Use & & \checkmark & none \\ 
\hline
\rowcolor{lightorange} \textbf{H5.4} & \textbf{Trust: Ability} & \textbf{decrease} & & \textbf{none} \\ 
\hline
\rowcolor{lightred} \textbf{H5.5} & \textbf{Trust: Benevolence} & \textbf{decrease} & & \textbf{increase} \\ 
\hline
\textbf{H5.6} & Trust: Integrity & & \checkmark & none \\ 
\hline
\rowcolor{lightorange} \textbf{H5.7} & \textbf{Trust: General} & \textbf{decrease} & & \textbf{none} \\ 
\hline
\rowcolor{lightorange} \textbf{H5.8} & \textbf{Attitude} & \textbf{decrease} & & \textbf{none} \\ 
\hline
\end{tabular}
\end{table}

\subsection*{Overview of Findings}

The results of the hypothesis testing went against most pre-registered hypotheses, which were based on the literature as well as the pilot, which had a total of 43 participants, 21 in the ``trustworthy AI'' group and 22 in the ``reliable AI'' group. For hypotheses \textbf{H1}, \textbf{H2}, which predicted that the trustworthy AI condition would lead to lower ratings of blameworthiness and accountability, we did not observe any effect and could therefore not reject the null hypothesis. For Hypotheses \textbf{H3}, and \textbf{H4}, which concerned confidence ratings and where no special prediction was issued, no effect was observed. Finally, for Hypothesis \textbf{H5}, predicting a lower total TAM score, we could not reject the null hypothesis as we did not observe any significant difference.

Moving to specific aspects of trustworthy AI's impact, the results indicate a significant effect of the label ``trustworthy AI'' on only two of the TAM variables: higher ratings of perceived ease of use and benevolence, both of which differ from the preregistration, but suggest that the conditions did affect some aspects of technological acceptance. The other hypotheses did not reach significance.   

For a comprehensive overview of these results, please refer to Table \ref{table:hypotheses}.

\subsection*{H1 - H4: Vignette Responses}
\begin{figure}[ht]
    \centering
    \includegraphics[width=\textwidth]{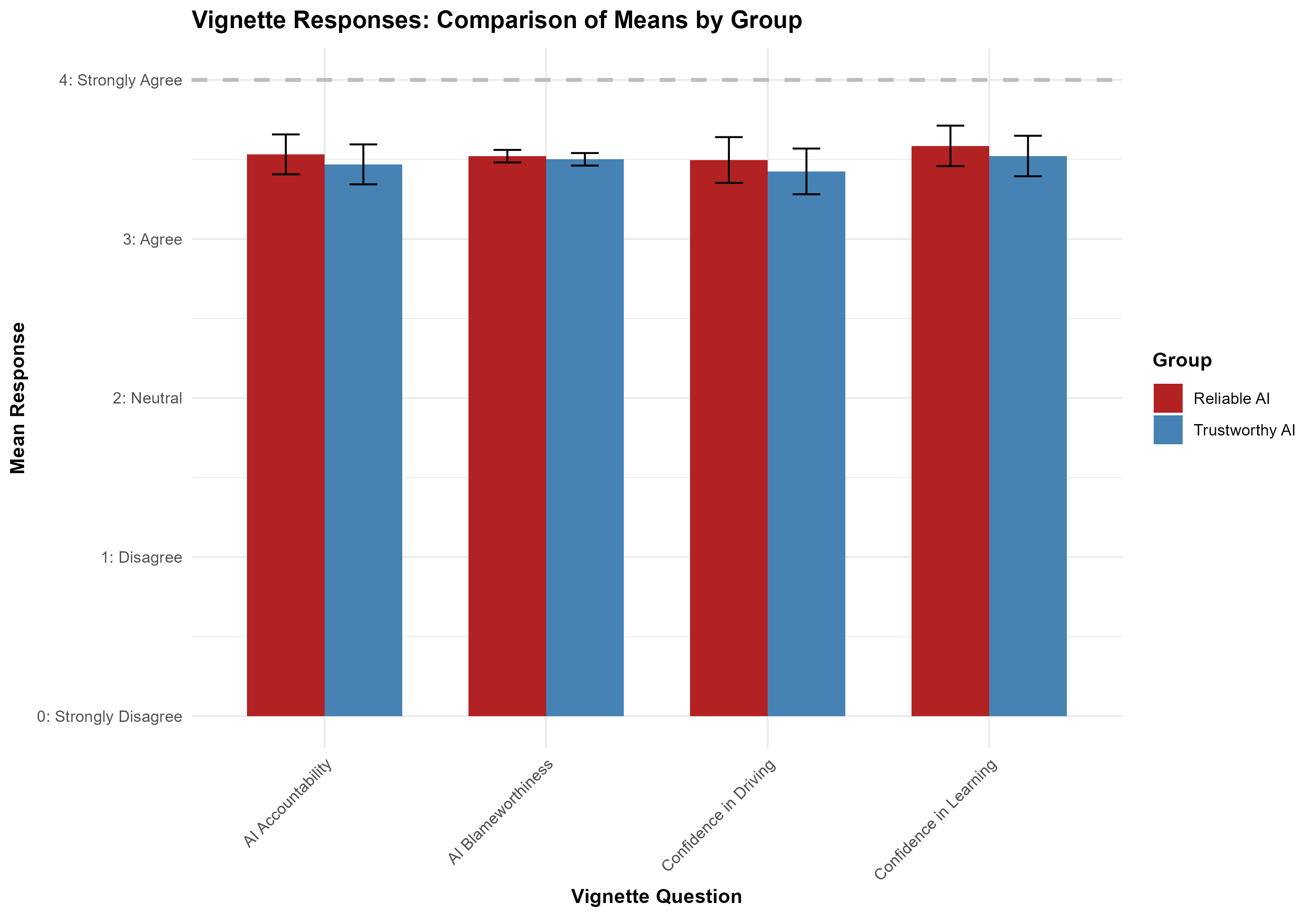}
    \caption{Results for H1 - H4: Vignette Responses.}
    \label{fig:Vignette_Responses}
\end{figure}

As detailed in the pre-registered plan, we conducted t-tests to compare group means on AI accountability, AI blameworthiness, confidence in driving, and confidence in learning between the trustworthy AI and reliable AI groups (See Figure \ref{fig:Vignette_Responses}). The analysis for each hypothesis showed that there were no significant differences between the trustworthy AI and reliable AI conditions across the measured variables. The \(p\)-values for all comparisons were greater than \(0.05\), indicating that the observed differences were not statistically significant, as confirmed  through Bayesian ordinal regression (See Table \ref{table:bayes}).


\subsubsection*{H1: AI Accountability}
There was no significant difference between the trustworthy AI group (mean = \(3.47\)) and the reliable AI group (mean = \(3.53\)) in terms of accountability, \(t(478) = 1.05\), \(p = 0.292\), with a 95\% confidence interval of \([-0.0541, 0.180]\).

\subsubsection*{H2: AI Blameworthiness}
The blameworthiness ratings were similar for the trustworthy AI group (mean = \(3.50\)) and the reliable AI group (mean = \(3.52\)), \(t(478) = 0.333\), \(p = 0.739\), with a 95\% confidence interval of \([-0.0963, 0.136]\).

\subsubsection*{H3: Confidence in Driving}
Participants' confidence in driving using the automotive AI did not significantly differ between the trustworthy AI group (mean = \(3.42\)) and the reliable AI group (mean = \(3.50\)), \(t(478) = 1.264\), \(p = 0.207\), with a 95\% confidence interval of \([-0.0398, 0.184]\).

\subsubsection*{H4: Confidence in Learning}
Similarly, confidence in learning to drive with the automotive AI was not significantly different between the trustworthy AI group (mean = \(3.52\)) and the reliable AI group (mean = \(3.59\)), \(t(478) = 1.07\), \(p = 0.285\), with a 95\% confidence interval of \([-0.0531, 0.180]\). \newline

\begin{figure}[ht]
    \centering
    \includegraphics[width=\textwidth]{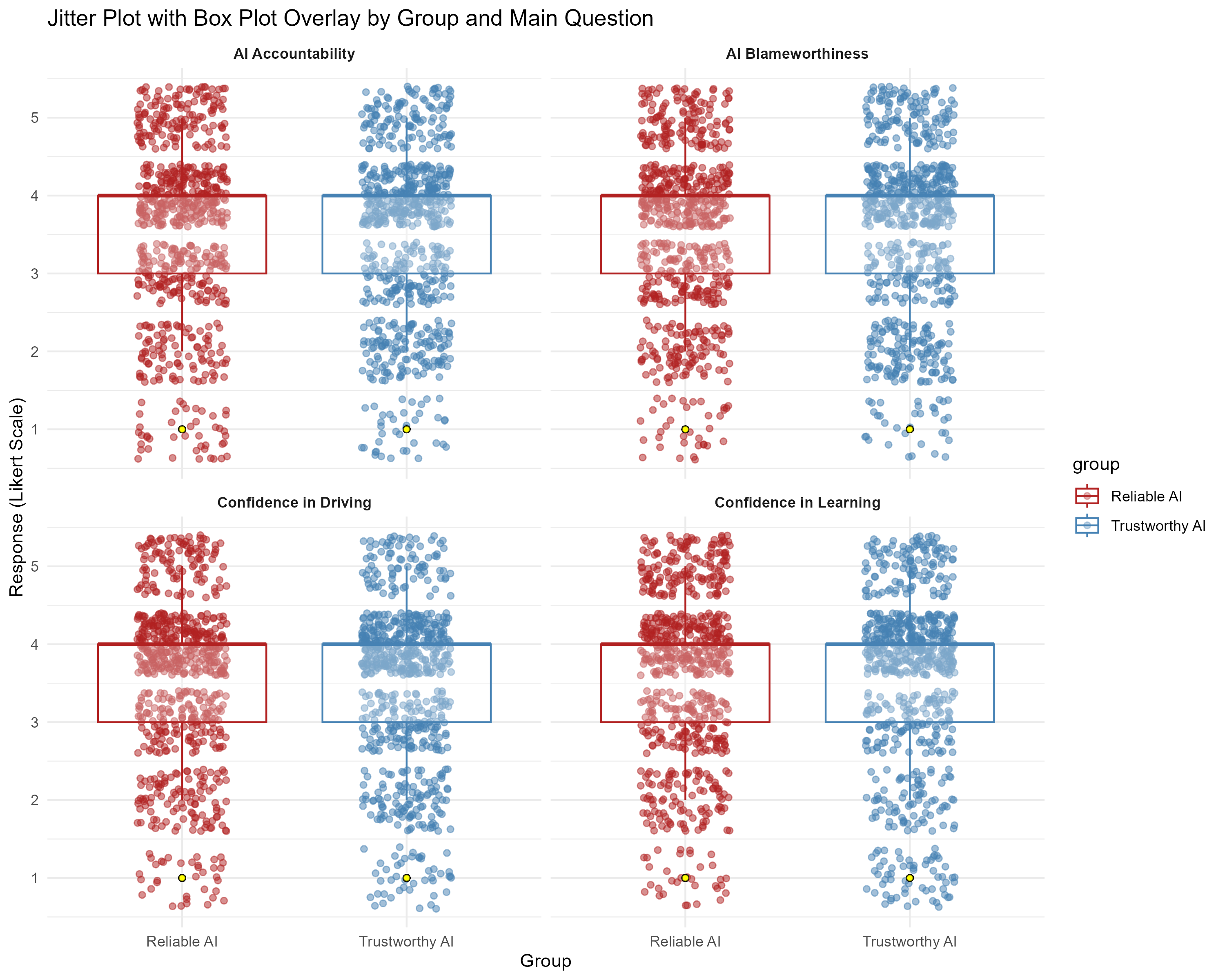}
    \caption{Jitter Plot with Box Plot Overlay for the Vignette Responses}
    \label{fig:Jitter_Box}
\end{figure}

The jitter plot with box plot overlay illustrates the distribution of responses across four key measures: AI Accountability, AI Blameworthiness, Confidence in Driving, and Confidence in Learning, comparing the ``Reliable AI'' and ``Trustworthy AI'' groups (See Figure \ref{fig:Jitter_Box}). The box plots, which summarize the central tendency and spread of the responses, show that the medians and interquartile ranges (IQRs) are similar across both conditions for each measure. The overlapping IQRs in the box plots indicate that participants in both conditions exhibit similar attitudes and perceptions, particularly in terms of confidence in driving and learning, as well as their views on AI accountability and blameworthiness.

To further examine the potential differences between the ``Reliable AI'' and ``Trustworthy AI'' conditions, Bayesian ordinal regression models were applied to each of the four questions in the vignette study (See Table \ref{table:bayes}). The analysis yielded small estimated group effects across all questions, with 95\% credible intervals consistently encompassing zero, indicating no strong evidence for significant differences between the groups. Specifically, the probabilities that the group effect is greater than zero were all below 0.5, further underscoring the absence of a positive effect. The Bayesian models demonstrated good convergence, with Rhat values close to 1.00 and sufficient effective sample sizes, reinforcing the reliability of the estimates (see \nameref{bayes_sup} in  \nameref{supplementary_material}). Overall, these results corroborate the earlier findings from the traditional t-tests and Wilcoxon tests, suggesting that any observed differences are minimal or non-existent.

\begin{table}[ht]
\centering
\caption{Summary of Bayesian Analysis Results for Vignette Questions}
\vspace{1em}
\label{table:bayes}
\begin{tabular}{|l|r|r|r|r|}
\hline
\textbf{Main Question} & 
\textbf{Estimate} & 
\makecell{\textbf{95\% CI} \\ \textbf{(Lower)}} & 
\makecell{\textbf{95\% CI} \\ \textbf{(Upper)}} & 
\makecell{\textbf{Probability} \\ \textbf{(Effect > 0)}} \\ 
\hline
AI Accountability             & -0.2175           & -0.8834                  & 0.4387                   & 0.26425                         \\ \hline
AI Blameworthiness            & -0.0296           & -0.6699                  & 0.6185                   & 0.46325                         \\ \hline
Confidence in Driving         & -0.3043           & -0.8843                  & 0.2742                   & 0.15650                         \\ \hline
Confidence in Learning        & -0.2941           & -0.9500                  & 0.3188                   & 0.17950                         \\ \hline
\end{tabular}
\end{table}

\subsection*{H5: Total TAM Score}

\begin{figure}[ht]
    \centering
    \includegraphics[width=\textwidth]{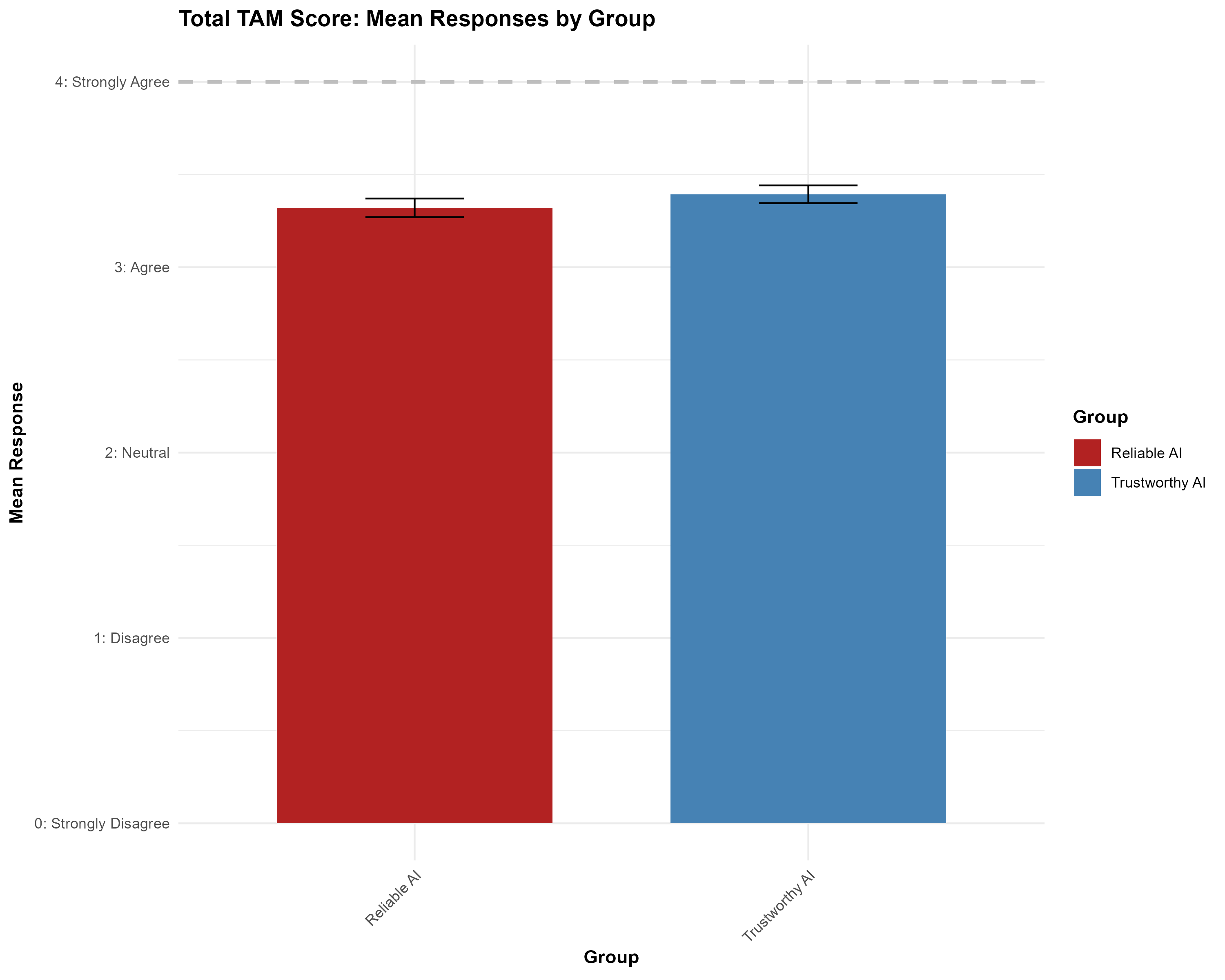}
    \caption{Results for Total TAM Score}
    \label{fig:TOTAL_TAM}
\end{figure}

For the Total TAM Score, we conducted ordinal regression analyses to examine the relationship between the condition and participants' perceptions across the total sum of the various questions. The mixed model had a log-likelihood of \(-4498.18\) and an AIC of \(9010.35\). The model converged after 538 iterations, with a maximum gradient of \(1.01 \times 10^{-3}\) and a condition number of \(1.8 \times 10^{02}\). The coefficient for the \texttt{grouptrust} variable was estimated at \(0.2058\) with a standard error of \(0.2137\), resulting in a \(z\)-value of \(0.963\) and a \(p\)-value of \(0.335\). The 95\% confidence interval for this estimate ranged from \(-0.2131\) to \(0.6247\), indicating no statistically significant effect of the ``trustworthy AI'' label on the Total TAM Score. The thresholds were estimated as follows: Strongly Disagree|Disagree at \(-4.0489\) (CI: \([-4.5804, -3.5173]\)), Disagree|Neutral at \(-2.2495\) (CI: \([-2.7636, -1.7354]\)), Neutral|Agree at \(0.0009\) (CI: \([-0.5069, 0.5086]\)), and Agree|Strongly Agree at \(3.4428\) (CI: \([2.9192, 3.9664]\)). None of the threshold differences reached statistical significance. The Brant test for the parallel regression assumption of the Cumulative Link Mixed Model (CLMM) indicated that the assumption holds, with an omnibus test result of \(\chi^2(3) = 7.11\), \(p = 0.07\). Specifically, the test for the \texttt{grouptrust} variable also yielded \(\chi^2(3) = 7.11\), \(p = 0.07\).

\subsection*{H5.1 - H5.8: Results of Individual Technology Acceptance Items}
\label{subsec:TAM}

\begin{figure}[H]
    \centering
    \includegraphics[width=\textwidth]{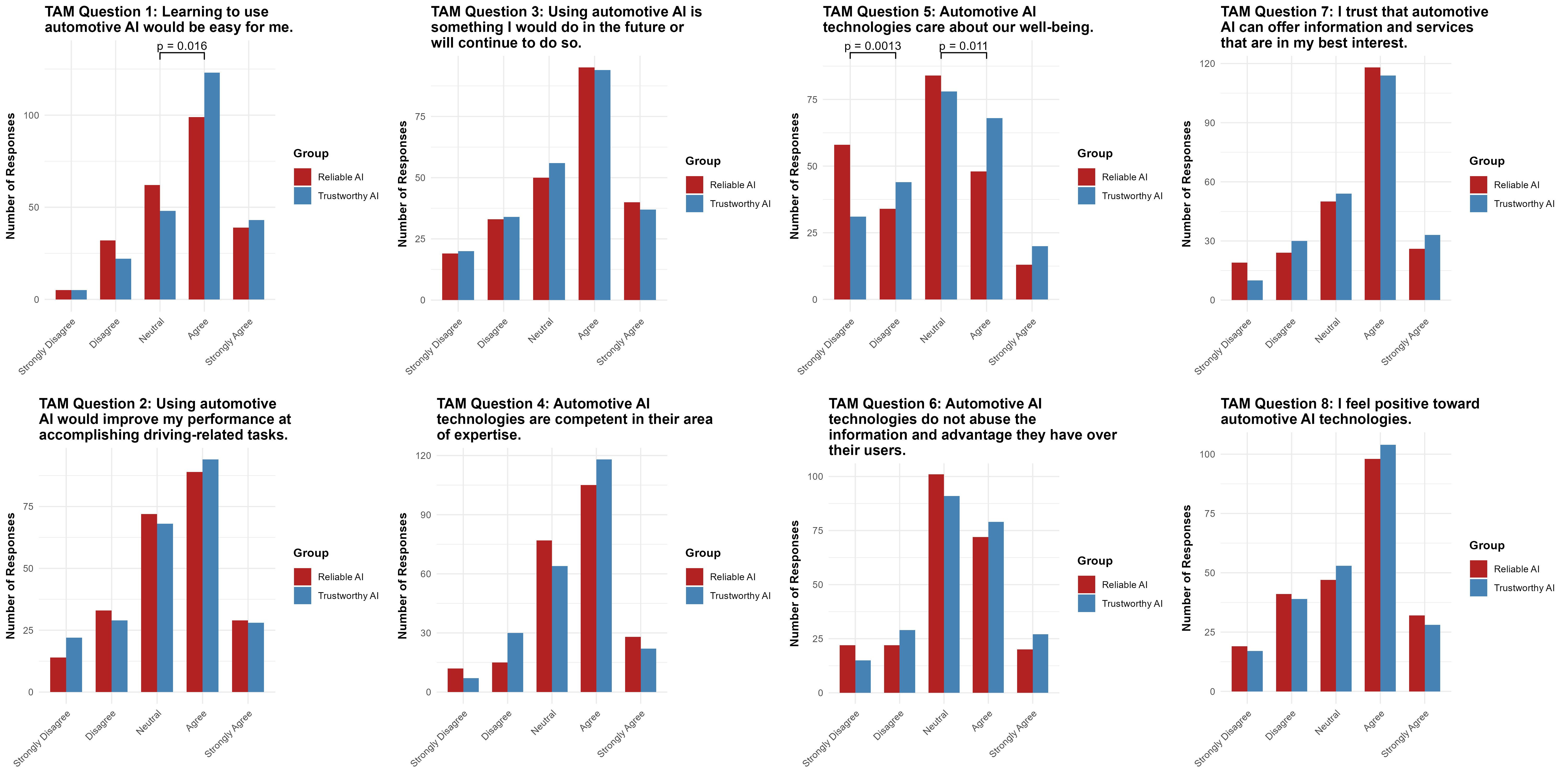}
    \caption{Results for H5.1 - H5.8: all eight TAM Questions.}
    \label{fig:Combined_TAM}
\end{figure}

For individual TAM items, we conducted ordinal regression analyses to examine the relationship between the group label and participants' perceptions across the various questions: 1. perceived ease of use, 2. perceived usefulness, 3. intention to use, 4. trust: ability, 5. trust: benevolence, 6. trust: integrity, 7. trust: general, 8. attitude general. As outlined in our pre-registration, we tested each question for adherence to the proportional odds assumption using the Brant test, implemented through the \texttt{brant} package in R. The results indicated that Question 4 (trust in ability) did not satisfy this assumption (\(\chi^2\) = \(8.65\), \(df = 3\), \(p = 0.0343)\). To maintain consistency across our analyses and ensure a uniform approach for all eight questions, we opted to use a model with flexible thresholds. This approach allowed us to account for potential variability in the relationship between the labels and responses across different thresholds. Each model included 478 observations, using a logit link function with flexible thresholds. Below are the detailed results for each question, including the specific model fit statistics, coefficients, and threshold estimates with their respective confidence intervals (see Figure \ref{fig:Combined_TAM}). 


\begin{figure}[H]
    \centering
    \includegraphics[width=\textwidth]{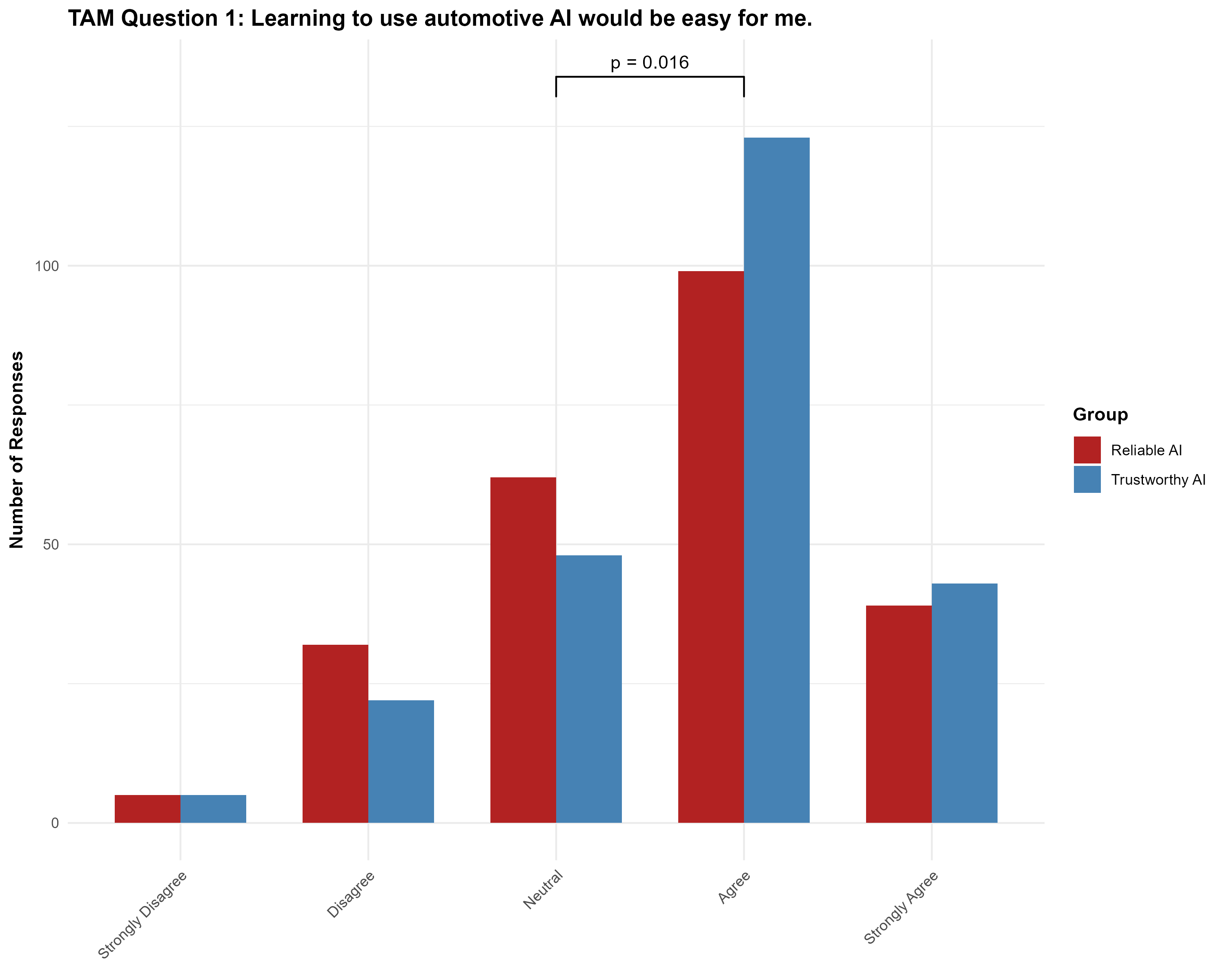}
    \caption{Results for TAM Question 1.}
    \label{fig:TAM_Question_1}
\end{figure}

\subsubsection*{Question 1: Learning to use automotive AI would be easy for me}

For Question 1, the model had a log-likelihood of \(-630.88\) and an AIC of \(1271.77\). The model converged after six iterations, with a maximum gradient of \(3.92 \times 10^{-13}\) and a condition number of \(2.5 \times 10^{01}\). The coefficient for the \texttt{grouptrust} variable was estimated at \(0.3354\) with a standard error of \(0.1697\), resulting in a \(z\)-value of \(1.976\) and a \(p\)-value of \(0.0481\). The 95\% confidence interval for this estimate ranged from \(0.0028\) to \(0.6679\), indicating a statistically significant effect of the ``trustworthy AI'' label on the perception of ease of learning. The thresholds were estimated as follows: Strongly Disagree|Disagree at \(-3.6908\) (CI: \([-4.3346, -3.0470]\)), Disagree|Neutral at \(-1.7084\) (CI: \([-2.0130, -1.4038]\)), Neutral|Agree at \(-0.3899\) (CI: \([-0.6386, -0.1412]\)), and Agree|Strongly Agree at \(1.7564\) (CI: \([1.4553, 2.0575]\)) (see Figure \ref{fig:TAM_Question_1}). The result for the Neutral|Agree threshold difference was significant (\(p = 0.016\)), with a standard error of \(0.1391\) and a difference of \(0.4624\). The other threshold differences did not reach significance.


\begin{figure}[H]
    \centering
    \includegraphics[width=\textwidth]{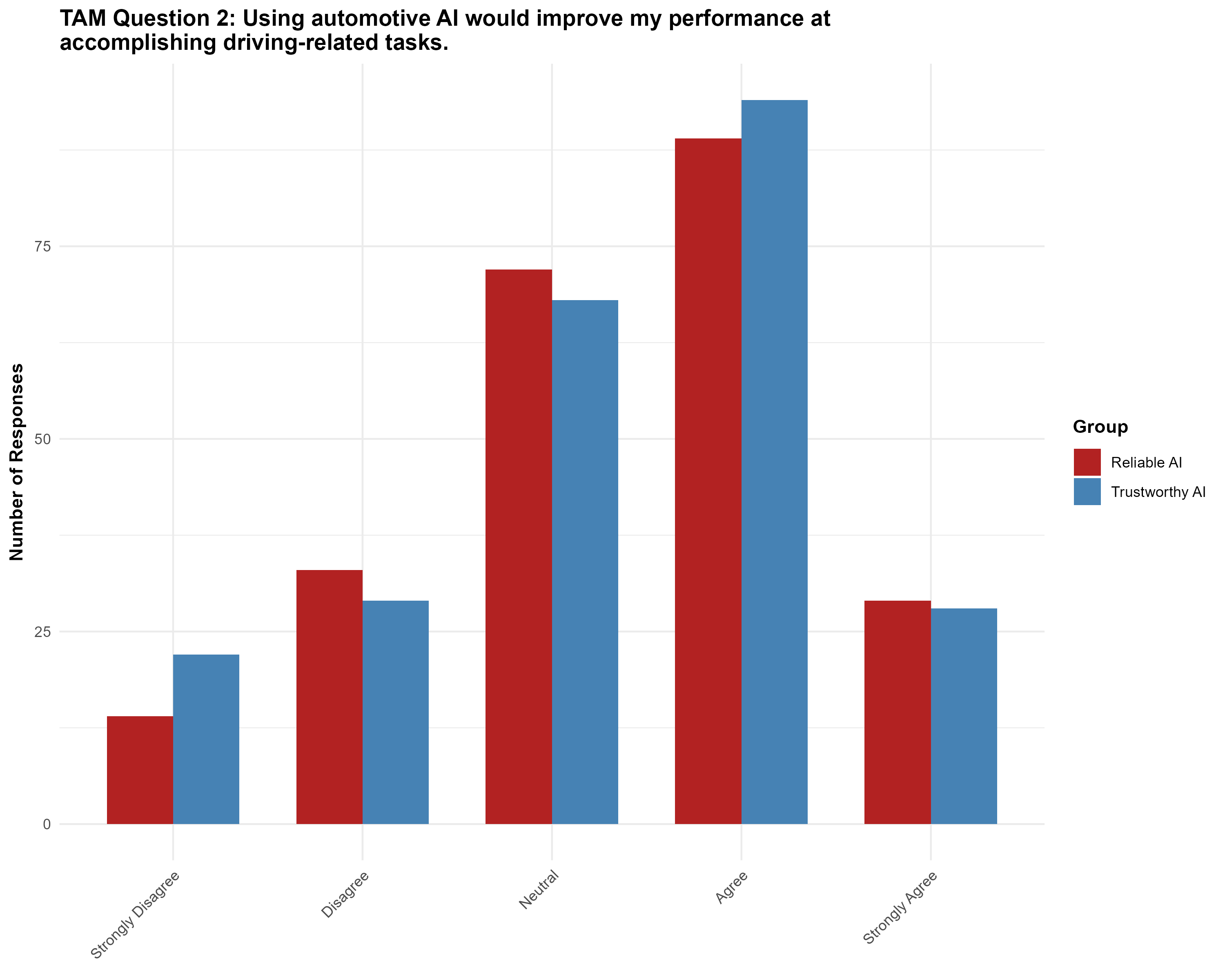}
    \caption{Results for TAM Question 2.}
    \label{fig:TAM_Question_2}
\end{figure}

\subsubsection*{Question 2: Using automotive AI would improve my performance at accomplishing driving-related tasks}

For Question 2, the model had a log-likelihood of \(-688.55\) and an AIC of \(1387.09\). The model converged after five iterations, with a maximum gradient of \(2.10 \times 10^{-08}\) and a condition number of \(1.9 \times 10^{01}\). The coefficient for the \texttt{grouptrust} variable was estimated at \(-0.0318\) with a standard error of \(0.1657\), resulting in a \(z\)-value of \(-0.192\) and a \(p\)-value of \(0.848\). The 95\% confidence interval for this estimate ranged from \(-0.3566\) to \(0.2929\), indicating that the effect of the ``trustworthy AI'' label on performance improvement is not statistically significant. The thresholds were estimated as follows: Strongly Disagree|Disagree at \(-2.5237\) (CI: \([-2.9003, -2.1470]\)), Disagree|Neutral at \(-1.3710\) (CI: \([-1.6465, -1.0956]\)), Neutral|Agree at \(-0.5171\) (CI: \([-0.7691, -0.2652]\)), and Agree|Strongly Agree at \(1.5071\) (CI: \([1.2077, 1.8065]\)). None of the thresholds reached statistical significance (see Figure \ref{fig:TAM_Question_2}).


\begin{figure}[H]
    \centering
    \includegraphics[width=\textwidth]{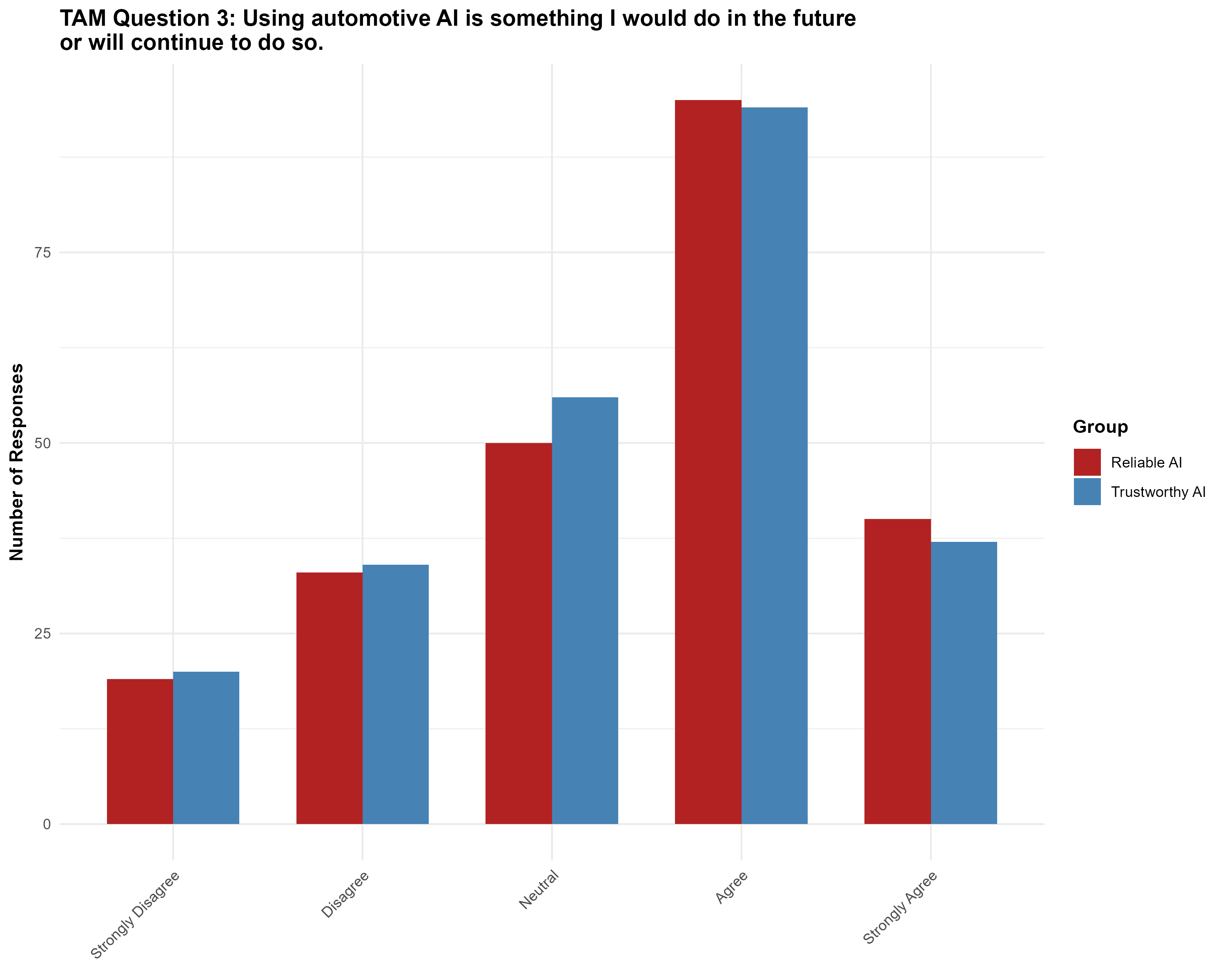}
    \caption{Results for TAM Question 3.}
    \label{fig:TAM_Question_3}
\end{figure}

\subsection*{H5.3: Intention to Use}
\label{subsec:H5.3}

We conducted an ordinal regression analysis to examine the relationship between the group labels, either ``trustworthy AI'' or ``reliable AI'' and the participants' intention to use the technology. The analysis used a logit link function with flexible thresholds. The model included 478 observations, with a log-likelihood of \(-704.85\) and an AIC of \(1419.70\). The model converged after five iterations, with a maximum gradient of \(7.07 \times 10^{-10}\) and a condition number of \(2.2 \times 10^{01}\).

The coefficient for the \texttt{grouptrust} variable was estimated at \(-0.0877\) with a standard error of \(0.1653\), resulting in a \(z\)-value of \(-0.531\) and a \(p\)-value of \(0.596\). The 95\% confidence interval for this estimate ranged from \(-0.4116\) to \(0.2363\), indicating that the effect of the ``trustworthy AI'' label on the intention to use is not statistically significant. This result supports the hypothesis, suggesting that the label ``trustworthy AI'' does not predict a change in the intention to use the technology (See Figure \ref{fig:TAM_Question_3}).

The threshold coefficients indicate the points at which the response transitions from one category to the next on the Likert scale. The threshold between ``Strongly Disagree'' and ``Disagree'' was estimated at \(-2.4660\) with a standard error of \(0.1878\), resulting in a \(z\)-value of \(-13.134\) and a 95\% confidence interval ranging from \(-2.8340\) to \(-2.0981\). The threshold between ``Disagree'' and ``Neutral'' was estimated at \(-1.3005\) with a standard error of \(0.1393\), resulting in a \(z\)-value of \(-9.336\) and a 95\% confidence interval ranging from \(-1.5735\) to \(-1.0275\). The threshold between ``Neutral'' and ``Agree'' was estimated at \(-0.2716\) with a standard error of \(0.1249\), resulting in a \(z\)-value of \(-2.175\) and a 95\% confidence interval ranging from \(-0.5163\) to \(-0.0268\). Finally, the threshold between ``Agree'' and ``Strongly Agree'' was estimated at \(1.6065\) with a standard error of \(0.1489\), resulting in a \(z\)-value of \(10.789\) and a 95\% confidence interval ranging from \(1.3147\) to \(1.8984\). In summary, the analysis indicates that the label ``trustworthy AI'' does not have a statistically significant effect on the intention to use the technology. 


\begin{figure}[H]
    \centering
    \includegraphics[width=\textwidth]{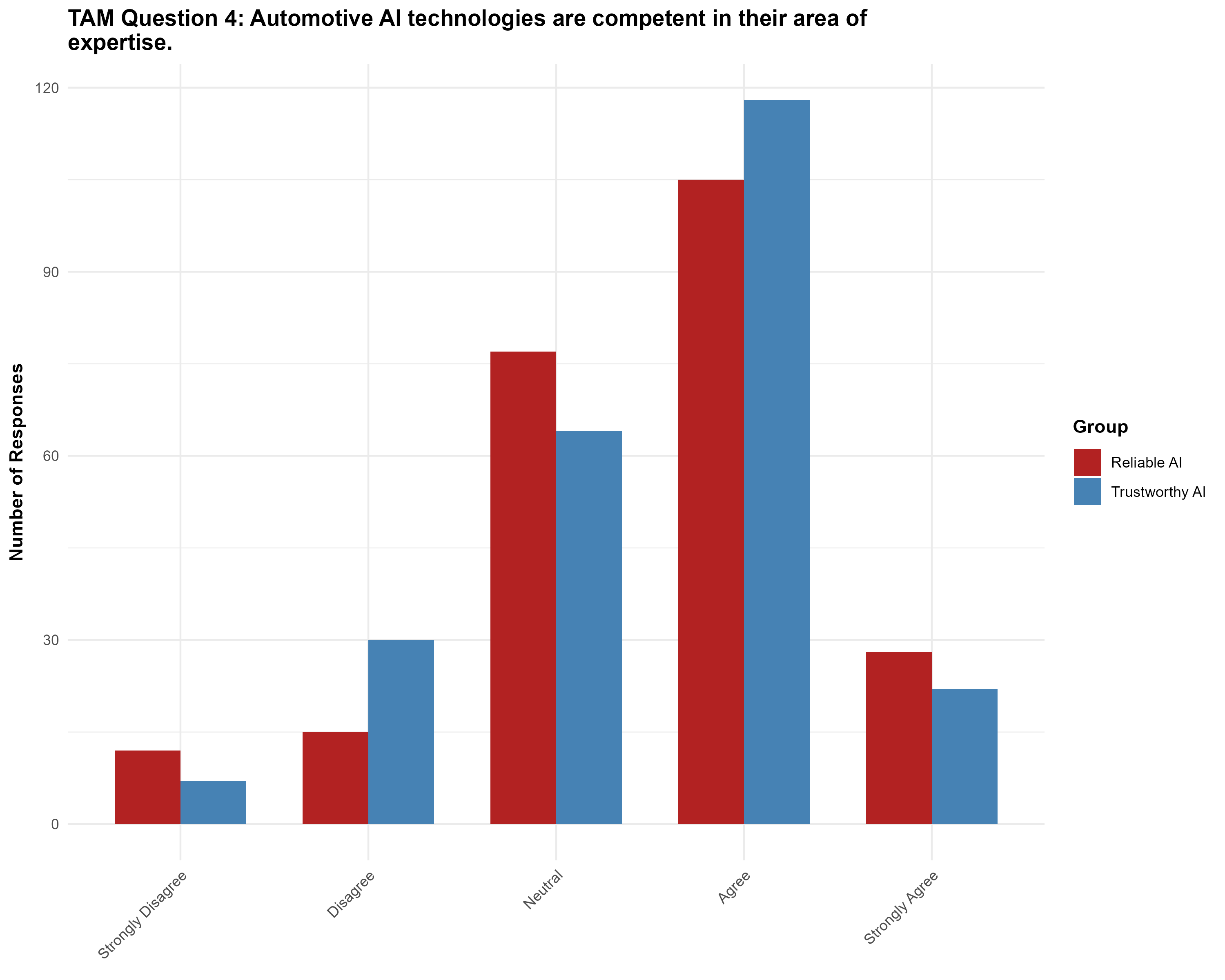}
    \caption{Results for TAM Question 4.}
    \label{fig:TAM_Question_4}
\end{figure}

\subsubsection*{Question 4: Automotive AI technologies are competent in their area of expertise}

For Question 4, the model had a log-likelihood of \(-622.62\) and an AIC of \(1255.24\). The model converged after six iterations, with a maximum gradient of \(3.13 \times 10^{-13}\) and a condition number of \(1.9 \times 10^{01}\). The coefficient for the \texttt{grouptrust} variable was estimated at \(-0.0442\) with a standard error of \(0.1698\), resulting in a \(z\)-value of \(-0.26\) and a \(p\)-value of \(0.795\). The 95\% confidence interval for this estimate ranged from \(-0.3771\) to \(0.2886\), indicating that the effect of the ``trustworthy AI'' label on the perception of competence is not statistically significant. The thresholds were estimated as follows: Strongly Disagree|Disagree at \(-3.2074\) (CI: \([-3.6975, -2.7173]\)), Disagree|Neutral at \(-1.8894\) (CI: \([-2.2027, -1.5762]\)), Neutral|Agree at \(-0.3087\) (CI: \([-0.5557, -0.0618]\)), and Agree|Strongly Agree at \(2.1247\) (CI: \([1.7871, 2.4623]\)) (see Figure \ref{fig:TAM_Question_4}).


\begin{figure}[H]
    \centering
    \includegraphics[width=\textwidth]{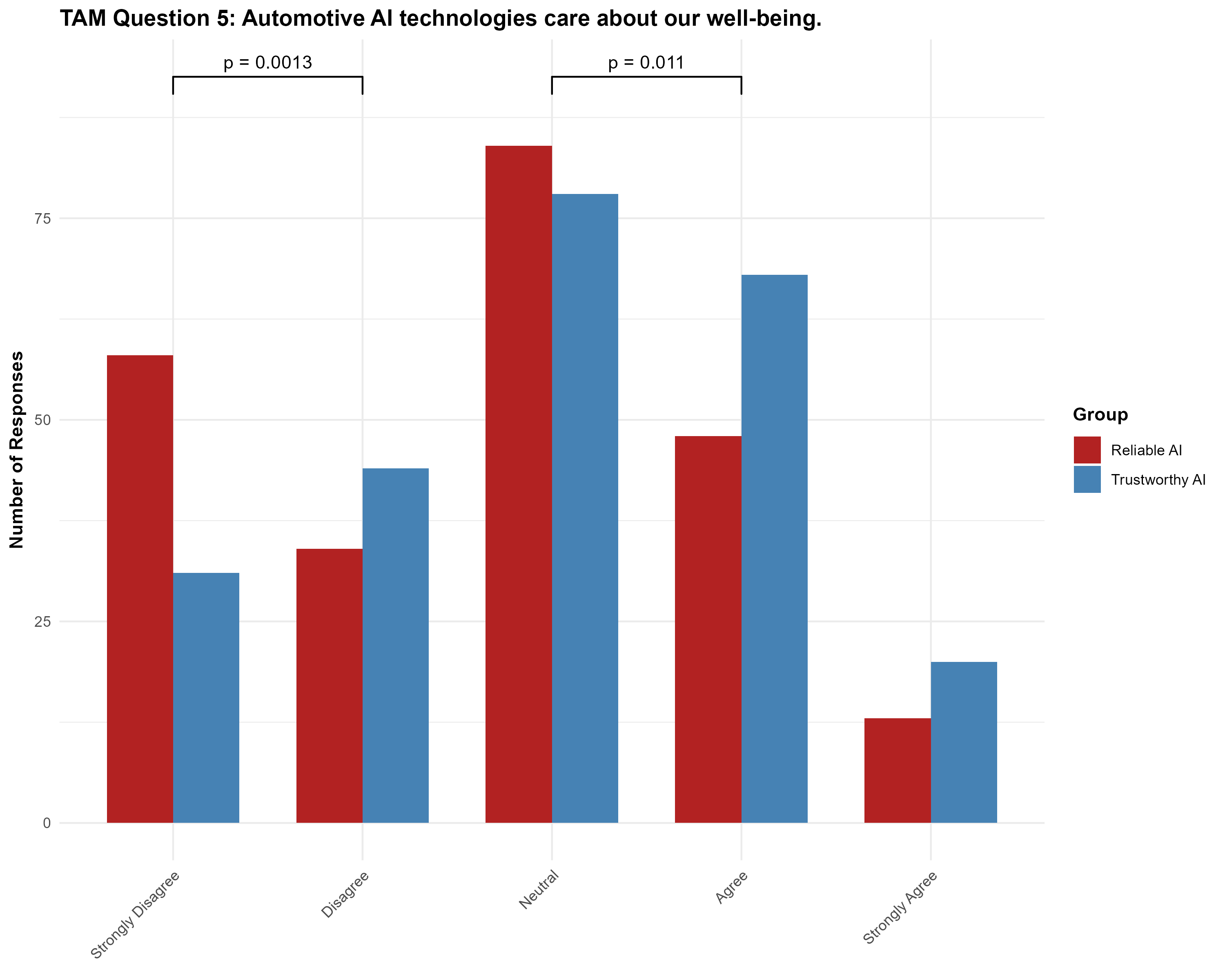}
    \caption{Results for TAM Question 5.}
    \label{fig:TAM_Question_5}
\end{figure}

\subsubsection*{Question 5: Automotive AI technologies care about our well-being}

For Question 5, the model had a log-likelihood of \(-714.40\) and an AIC of \(1438.81\). The model converged after five iterations, with a maximum gradient of \(8.39 \times 10^{-09}\) and a condition number of \(2.4 \times 10^{01}\). The coefficient for the \texttt{grouptrust} variable was estimated at \(0.4862\) with a standard error of \(0.1651\), resulting in a \(z\)-value of \(2.945\) and a \(p\)-value of \(0.00323\). The 95\% confidence interval for this estimate ranged from \(0.1626\) to \(0.8097\), indicating a statistically significant positive effect of the ``trustworthy AI'' label on the perception that automotive AI technologies care about our well-being. The thresholds were estimated as follows: Strongly Disagree|Disagree at \(-1.2419\) (CI: \([-1.5177, -0.9662]\)), Disagree|Neutral at \(-0.3804\) (CI: \([-0.6267, -0.1341]\)), Neutral|Agree at \(1.0483\) (CI: \([0.7878, 1.3087]\)), and Agree|Strongly Agree at \(2.8723\) (CI: \([2.4721, 3.2725]\)) (see Figure \ref{fig:TAM_Question_5}). The results for the Strongly Disagree\(|\)Disagree threshold difference was significant (\(p = 0.0013\)), with a standard error of \(0.1924\) and a difference of \(0.7862\). The result for the Neutral\(|\)Agree threshold difference was also significant (\(p = 0.011\)), with a standard error of \(0.1338\) and a difference of \(0.5065\). The other threshold differences did not reach significance.


\begin{figure}[H]
    \centering
    \includegraphics[width=\textwidth]{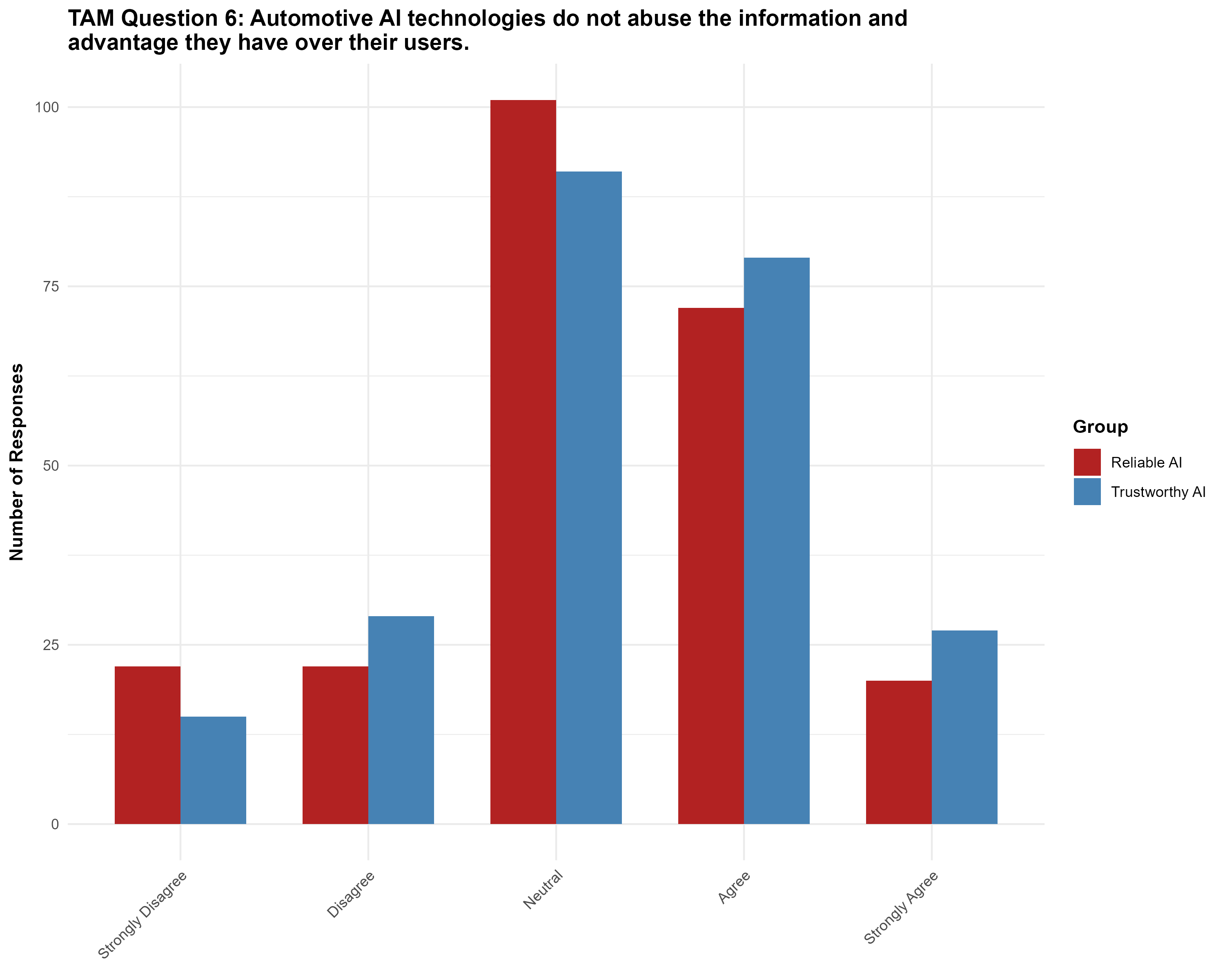}
    \caption{Results for TAM Question 6.}
    \label{fig:TAM_Question_6}
\end{figure}

\subsubsection*{Question 6: Automotive AI technologies do not abuse the information and advantage they have over their users}

For Question 6, the model had a log-likelihood of \(-666.32\) and an AIC of \(1342.65\). The model converged after five iterations, with a maximum gradient of \(1.14 \times 10^{-07}\) and a condition number of \(1.6 \times 10^{01}\). The coefficient for the \texttt{grouptrust} variable was estimated at \(0.1857\) with a standard error of \(0.1671\), resulting in a \(z\)-value of \(1.111\) and a \(p\)-value of \(0.267\). The 95\% confidence interval for this estimate ranged from \(-0.1419\) to \(0.5132\), indicating that the effect of the ``trustworthy AI'' label on the perception of information abuse is not statistically significant. The thresholds were estimated as follows: Strongly Disagree|Disagree at \(-2.3873\) (CI: \([-2.7580, -2.0166]\)), Disagree|Neutral at \(-1.3978\) (CI: \([-1.6782, -1.1173]\)), Neutral|Agree at \(0.4396\) (CI: \([0.1940, 0.6852]\)), and Agree|Strongly Agree at \(2.3124\) (CI: \([1.9653, 2.6596]\)). None of the thresholds reached statistical significance (see Figure \ref{fig:TAM_Question_6}).


\begin{figure}[H]
    \centering
    \includegraphics[width=\textwidth]{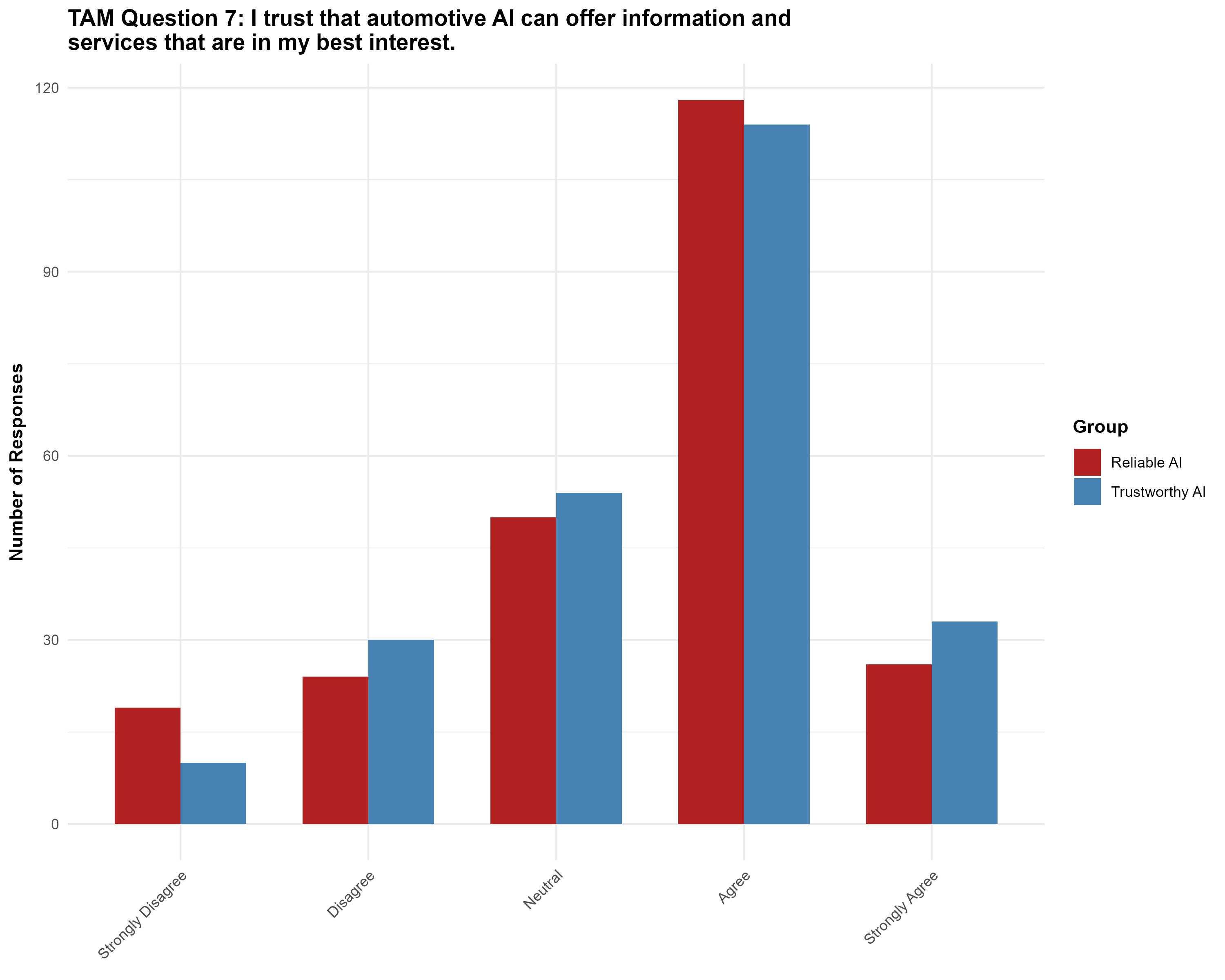}
    \caption{Results for TAM Question 7.}
    \label{fig:TAM_Question_7}
\end{figure}

\subsubsection*{Question 7: I trust that automotive AI can offer information and service that is in my best interest}

For Question 7, the model had a log-likelihood of \(-648.60\) and an AIC of \(1307.19\). The model converged after six iterations, with a maximum gradient of \(5.36 \times 10^{-14}\) and a condition number of \(2.0 \times 10^{01}\). The coefficient for the \texttt{grouptrust} variable was estimated at \(0.1035\) with a standard error of \(0.1698\), resulting in a \(z\)-value of \(0.61\) and a \(p\)-value of \(0.542\). The 95\% confidence interval for this estimate ranged from \(-0.2292\) to \(0.4362\), indicating that the effect of the ``trustworthy AI'' label on the perception that automotive AI offers information and service in the user's best interest is not statistically significant, and none of the thresholds reached statistical significance (see Figure \ref{fig:TAM_Question_7}).


\begin{figure}[H]
    \centering
    \includegraphics[width=\textwidth]{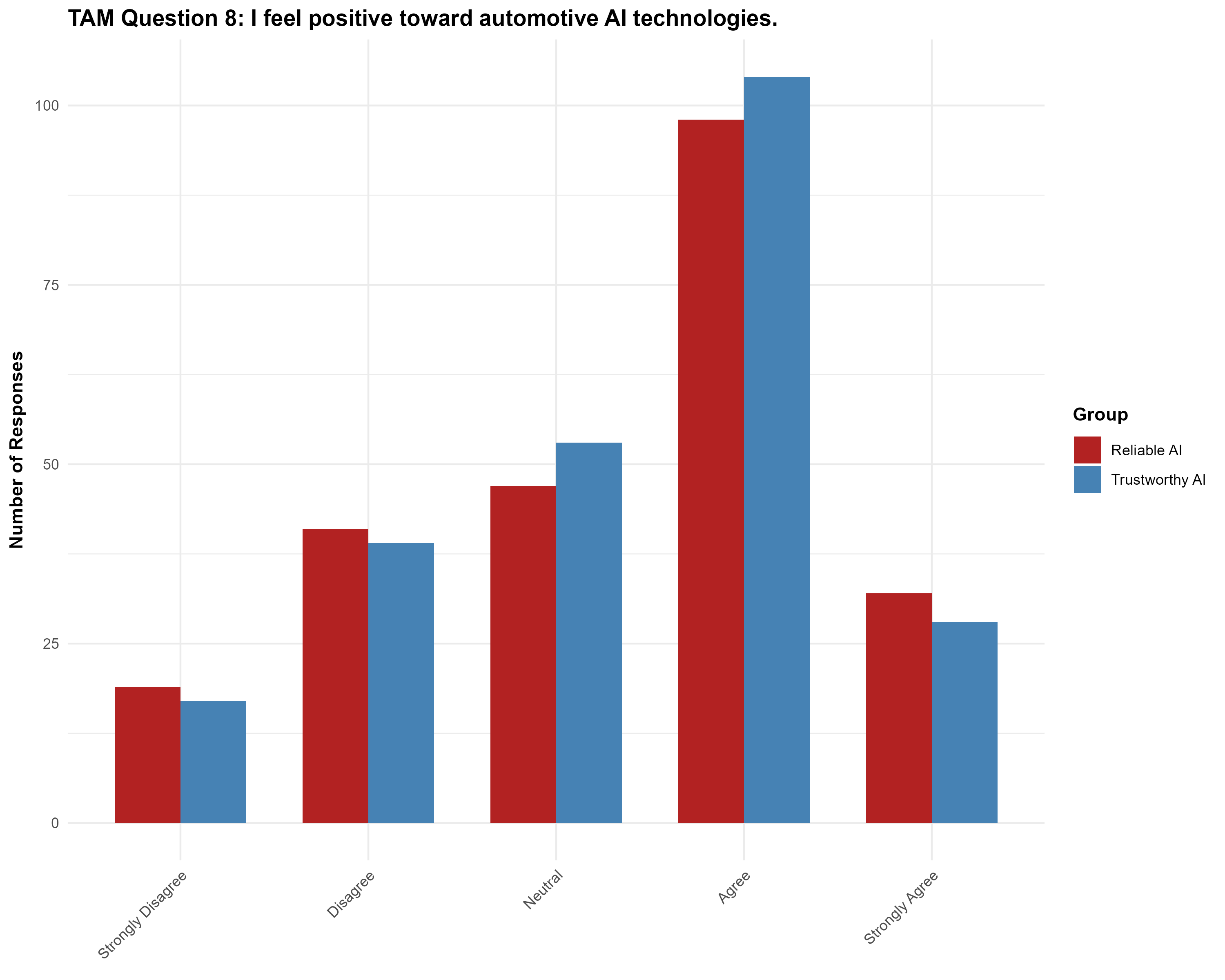}
    \caption{Results for TAM Question 8.}
    \label{fig:TAM_Question_8}
\end{figure}

\subsubsection*{Question 8: I feel positive toward automotive AI technologies}

For Question 8, the model had a log-likelihood of \(-691.06\) and an AIC of \(1392.11\). The model converged after five iterations, with a maximum gradient of \(1.10 \times 10^{-07}\) and a condition number of \(2.4 \times 10^{01}\). The coefficient for the \texttt{grouptrust} variable was estimated at \(-0.0022\) with a standard error of \(0.1663\), resulting in a \(z\)-value of \(-0.013\) and a \(p\)-value of \(0.989\). The 95\% confidence interval for this estimate ranged from \(-0.3282\) to \(0.3238\), indicating that the effect of the ``trustworthy AI'' label on the positive perception of automotive AI technologies is not statistically significant. No threshold differences reached significance (see Figure \ref{fig:TAM_Question_8}).

\section*{Discussion}

This study aimed to investigate the impact of labeling AI as ``trustworthy'' versus ``reliable'' on user perceptions and acceptance of automotive AI technologies. Regarding specific scenarios, the study found no significant differences between the trustworthy AI and reliable AI conditions. Users were neither more lenient nor more judgmental when AI was labeled as ``trustworthy'' or ``reliable'' in terms of accountability and blameworthiness. The same held for confidence in using and learning to use the technology. Although these findings go opposite to the prediction, they also suggest that nothing is gained especially in terms of confidence in using an anthropomorphic "trustworthy" label.  

The other focus of the study was on the Technology Acceptance Model The exploration of Total TAM score (\textbf{H5}) and sub-categories (\textbf{H5.1 - H5.8}) revealed mixed results. There was no difference in total scores, which aligns with some existing results which found that anthropomorphisation of AI did not affect trust attitudes (\cite{bib21}). Ultimately, the results suggest that while the trustworthiness label may enhance perceived ease of use and the perception that AI cares about people's wellbeing, it does not affect other dimensions. Still, the label ``trustworthy'' appears to cause users to anthropomorphize the AI, attributing more human-like qualities like benevolence to the system.

\subsubsection*{Ethical AI Design}

Given that the trustworthy label did not lead to higher acceptance or confidence, it is more practical and ethically sound to communicate the reliability of these systems. This aligns with the proposal that machines, being performance-based entities, are better suited to meet reliability standards, which can be objectively measured and verified, rather than trustworthiness, which involves normative judgments.


\subsubsection*{Practical Implications}
For developers and policymakers in the automotive AI industry, the study's findings suggest that there is no loss in using reliability in their communication practices. This label also ensures that AI systems are not described in a manner that will mislead users about the capabilities and limitations of AI systems.

\subsection*{Limitations}
This study has several limitations that need to be acknowledged. First, the observed increase in benevolence could be attributed to a framing effect due to the specific wording in the definition of trustworthy AI provided to the participants in the \texttt{grouptrust} condition (see \nameref{defs} for more details). The description highlighted that trustworthy AI ``should care about ethical norms and/or conceptualize moral principles'', which might have influenced participants' perceptions. Given the complex nature of trust and the various philosophical debates surrounding its definition, future research will need to walk a difficult tightrope. 


Second, while we observed that increased perceived usefulness and benevolence did not lead to a corresponding increase in intention to use, contrary to what would have been expected (see \cite{bib12, bib13, bib14}), this may be due to other factors unrelated to these variables. For instance, it is possible that participants had already formed opinions about whether they would use this technology, which were not influenced by the labels provided in the study. Future research could address this by employing a within-subject design to assess whether changing the label impacts participants' attitudes over the course of the study.

Third, this study only measured user attitudes toward \textit{automotive} AI. Future research should consider the effect of the trustworthy label on other AI technologies that are less tool-like, such as care-bots and chatbots.


\section*{Conclusion}


Labeling AI as ``trustworthy'' rather than "reliable" does not significantly impact user attitudes or acceptance, except increasing the likelihood that users will perceive AI as caring about our wellbeing. The findings on the lack of effect on users should be distinguished from the philosophical and ethical arguments on the lack of justification for attribution of trust to AI. If emphasizing reliability over trustworthiness is more practical and ethically sound in the development and deployment of AI, the findings suggest that there is no cost to do so in future policies and developer practices in the AI industry. 

\section*{Author Contributions}
\textbf{John Dorsch:} Conceptualization, Methodology, Software, Validation, Formal Analysis, Investigation, Resources, Data Curation, Writing - Original Draft, Writing - Review \& Editing, Visualization, Project administration. \textbf{Ophelia Deroy:} Conceptualization, Methodology, Validation, Resources, Writing - Review \& Editing, Supervision, Project administration, Funding acquisition.

\section*{Acknowledgments}
This work was supported by Bayerisches Forschungsinstitut für Digitale Transformation (bidt) Co-Learn Award Number KON-21-0000048.


%
%
%

\appendix
\section*{Supplementary Material}
\label{supplementary_material}

\subsection*{Goodness of Fit Evaluation}

\subsubsection*{T-Test Analysis and Further Analysis: Vignette Questions}
\label{T-Test_Sup}

The Shapiro-Wilk test results for the residuals revealed significant deviations from normality across all questions. For ``AI Accountability,'' the Shapiro-Wilk statistic was \(0.898\) with a \(p\)-value of \(8.42 \times 10^{-30}\). Similarly, ``AI Blameworthiness'' had a statistic of \(0.890\) and a \(p\)-value of \(9.94 \times 10^{-31}\), and ``Confidence in Driving'' showed a statistic of \(0.890\) with a \(p\)-value of \(8.59 \times 10^{-31}\). For ``Confidence in Learning,'' the Shapiro-Wilk statistic was \(0.884\) with a \(p\)-value of \(1.70 \times 10^{-31}\). These results consistently demonstrate that the residuals significantly deviate from a normal distribution, which invalidates the assumptions underlying the t-tests that were initially planned for the analysis.

Given the violation of normality assumptions in the residuals, the Wilcoxon test was employed as a non-parametric alternative to assess differences between the groups. This test is particularly suitable because it does not rely on the assumption of normality and is robust to deviations from this assumption. The results of the Mann-Whitney U tests showed no significant differences between the groups for any of the main questions. For ``AI Accountability,'' the Wilcoxon statistic was 265,040 with a \(p\)-value of \(0.286\). ``AI Blameworthiness'' had a Wilcoxon statistic of \(258,296\) and a \(p\)-value of \(0.866\). For ``Confidence in Driving,'' the statistic was \(265,950\) with a \(p\)-value of \(0.229\), and for ``Confidence in Learning,'' the statistic was 262,538 with a \(p\)-value of \(0.460\). These results confirm that, even when using a test suited for non-normal data, there are no significant differences between the groups for the questions examined.

\subsubsection*{CLMM Analysis: Total TAM Score}

Given that the Cumulative Link Mixed Model does not assume a normal distribution of residuals, we assessed the goodness of fit using likelihood ratio tests instead of residual analysis. The likelihood ratio test compares the fit of a null model, which includes only the random effects, to a full model that also includes the group effect. The results of this test indicate that the inclusion of the group variable does not significantly improve the model fit. Specifically, the likelihood ratio statistic was \(0.9269\) with a \(p\)-value of \(0.3357\), suggesting that the model's fit does not improve significantly when the group variable is included. This supports the finding that the group effect is not statistically significant, as reflected in the earlier analysis. The non-significant \(p\)-value further confirms that the inclusion of the group variable does not contribute meaningfully to explaining the variability in the response.

\subsubsection*{CLM Analysis: Individual TAM Questions}

For the same reason as above, we have evaluated the goodness of fit of Cumulative Link Models through likelihood ratio tests, instead of residual analysis as originally planned. These tests compare the fit of the base model, which includes only the group variable, to models incorporating additional predictors. For the base model in each TAM question, the likelihood ratio test results reveal that, in all but one case, adding additional predictors does not significantly improve model fit. Specifically, \(p\)-values for Questions 1, 2, 3, 6, 7, and 8 range from 0.245 to 0.980, indicating that the base model performs comparably to models with added predictors. The exception is Question 4, where a \(p\)-value of 0.029 suggests a marginal improvement in model fit with additional predictors. Thus, overall, the base model, which includes only the group variable, appears to be a generally adequate fit for the data across questions. For example, none of the models showed a significant improvement in fit when including gender as a predictor and its interaction with group (\(p\)-values ranged from 0.387 to 0.896), except for Question 6 where the \(p\)-value was 0.013, indicating an improvement. However, this did not affect the significance of the findings. For instance, the \(p\)-value for the coefficient of \texttt{grouptrust} is 0.757, for \texttt{Female} is 0.163, and for the interaction term \texttt{grouptrust} is 0.426.

\subsection*{Outlier Analysis}

We conducted an outlier analysis using the Interquartile Range (IQR) method to identify extreme values across each TAM question. Specifically, outliers were defined as observations that fell below the first quartile (Q1) minus $1.5 \times$ the IQR or above the third quartile (Q3) plus $1.5 \times$ the IQR.

Across all eight questions, the total number of responses was consistent at 478, with the majority of outliers selecting ``Strongly Disagree'', the one exception being Question 5, which had no outliers. For Question 5, the median response was ``Agree,'' with an interquartile range (IQR) from ``Neutral'' to ``Agree.''

For Question 1, there were 10 outliers, with a median of ``Neutral'' and an IQR from ``Disagree'' to ``Agree.'' Question 2 had 36 outliers, with a median of Neutral'' and an IQR of Disagree'' to ``Agree.'' In Question 3, 39 outliers were identified, with a median of Disagree'' and an IQR from ``Disagree'' to ``Agree.'' Question 4 showed 19 outliers, with a median of ``Disagree'' and an IQR from ``Disagree'' to ``Agree.'' Question 6 had 37 outliers, with a median of Neutral'' and an IQR from ``Disagree'' to ``Agree.'' For Question 7, 29 outliers were noted, with a median of Disagree'' and an IQR from ``Disagree'' to ``Agree.'' Finally, Question 8 had 36 outliers, with a median of ``Neutral'' and an IQR from ``Disagree'' to ``Agree.''

Consistent with our pre-registered plan, these outliers were not excluded from analysis. The rationale for retaining outliers is based on the premise that they may reflect genuine variability in participant responses. Excluding these data points could potentially bias our results and lead to an underestimation of the true variability within the sample. Therefore, all identified outliers were included in the final analyses to ensure a comprehensive representation of participant behavior and perceptions. In our analysis, the presence of outliers did not significantly distort the model estimates or alter the interpretation of the results. This supports our decision to retain these data points, as they provide valuable insights into the behavior and perceptions of individuals at the extremes of the distribution.

\subsection*{Bayesian Analysis of Vignette Questions}
\label{bayes_sup}

\subsubsection*{Model Summary}

The Bayesian cumulative link mixed model (CLMM) was fitted using the following settings:
\footnotesize
\begin{verbatim}
Family: cumulative 
Links: mu = logit; disc = identity 
Formula: response ~ group + (1 | participant) + (1 | vignette) 
Data: current_data (Number of observations: 1434) 
Draws: 4 chains, each with iter = 4000; warmup = 1000; thin = 1;
total post-warmup draws = 12000
\end{verbatim}
\normalsize

\subsubsection*{Multilevel Hyperparameters}

The multilevel (random effects) parameters for participants and vignettes are as follows:
\footnotesize
\begin{verbatim}
~participant (Number of levels: 478) 
Estimate    Est.Error    l-95% CI    u-95% CI    Rhat    Bulk_ESS    Tail_ESS
3.31        0.18         2.96        3.68        1.00    2740        4853

~vignette (Number of levels: 3) 
Estimate    Est.Error    l-95% CI    u-95% CI    Rhat    Bulk_ESS    Tail_ESS
0.53        0.55         0.08        2.06        1.00    2942        4237
\end{verbatim}
\normalsize

\subsubsection*{Regression Coefficients}

The regression coefficients for the model, including the intercepts and the group effect, are provided below:
\scriptsize
\begin{verbatim}
Regression Coefficients:
Estimate       Estimate   Est. Error  l-95% CI    u-95% CI   Rhat    Bulk_ESS    Tail_ESS
Intercept[1]   -5.71        0.49        -6.65       -4.67    1.00       2586       3749
Intercept[2]   -3.23        0.45        -4.09       -2.26    1.00       2239       3712
Intercept[3]   -1.14        0.44        -1.96       -0.18    1.00       2273       3740
Intercept[4]    3.13        0.45         2.33        4.12    1.00       2710       3530
groupt         -0.29        0.32        -0.95        0.32    1.00       1741       3197
\end{verbatim}
\normalsize

\subsubsection*{Conclusion}

The Bayesian analysis results indicate that there is no strong evidence of a significant difference between the ``Reliable AI'' and ``Trustworthy AI'' groups across the main questions. The group effect estimates were close to zero, with the 95\% credible intervals including zero, and the probabilities that the group effect is greater than zero were consistently below 0.5.

\subsubsection*{Descriptive Statistics}
\label{demo}
The demographic characteristics of the sample are presented below. The sample consisted of 493 participants with a balanced distribution in terms of gender and age groups. 15 participants were excluded from analysis due to being 65 years old or older, leaving a final sample of 478 participants whose data were analyzed. The majority of participants were either male (246) or female (242). There were also 3 participants who identified as non-binary, 1 participant who preferred not to say, and 1 who identified as `other'. The age distribution was as follows: 68 participants were 18-30 years old, 240 were 30-45 years old, 167 were 45-64 years old, 15 were 65 years or older, and 3 preferred not to say. A vast majority of participants held a driver's license (476), while 14 participants did not, and 3 preferred not to disclose their status.

The participants reported various areas of expertise, including 9 in philosophy, 32 in computer science, 5 in informatics, 8 in data science, 10 in computer programming, 3 in machine learning, and 1 in robotics, with no participants selecting `other'. The educational background of participants varied widely: 212 had a bachelor’s degree (4-year), 53 had an associate degree (2-year), 52 had a master’s degree, 81 had some college but no degree, 8 had a professional degree (JD, MD), 68 had completed high school (including GED), 3 had less than high school education, 8 had technical certification, 7 had a doctoral degree, and 1 preferred not to say. No participants selected `other' for education. Participants had various experiences with autonomous technology, with 221 participants reporting some experience with automotive AI, such as with lane-keeping, steering, route-planning, or parking assistance (see \nameref{exp} for the specific questions).

\subsection*{Definitions of Key Terms}
\label{defs}

Participants were instructed to read and understand the definitions of ``trustworthy'' and ``reliable'' AI, as appropriate to their group assignment. These definitions were also reiterated when participants read the vignettes. Definitions were developed and informed by normative analysis and ethical considerations in previous work in the ethics of AI \cite{bib3}.

\subsubsection*{Trustworthy AI}

\begin{enumerate}
    \item It should be lawful, complying with all applicable laws and regulations.
    \item It should be ethical and do the right thing for the right ethical reasons.
    \item It should care about ethical norms and/or conceptualize moral principles.
\end{enumerate}

\subsubsection*{Reliable AI}

\begin{enumerate}
    \item It should be lawful, complying with all applicable laws and regulations.
    \item It should be effective, shown to produce the expected outcome for the tasks it was assigned.
    \item It should be consistent, likely to produce such outcomes in the future.
\end{enumerate}

\subsection*{Inclusion Criteria}
\label{inc}

To be considered for inclusion in the study, participants were required to demonstrate a minimum threshold of fluency in English. Specifically, only those achieving a success rate of at least 75\% on the language check were included in the final analysis. Additionally, participants were required to meet a minimum accuracy threshold of 75\% on the attention check to ensure comprehension of the key components, one that was specific to group assignment.

\subsubsection*{Language Check}
\label{lang}
Please select \textbf{all} the sentences below that are written in \textbf{correct English}:

\begin{enumerate}
    \item Atheletes often need to warm up.
    \item I just saw a moose running down the road!
    \item John ill felt and went doctor.
    \item Could you the books put in that boxes?
    \item I am forget do my homeworks.
    \item She think English is more easier to learn.
    \item Tommorrow was tatiehr than today.
    \item The building is a very murnnlye.
    \item Where's the pen I gave you yesterday?
    \item He was pulled over by the police for driving 120 miles per hour.
\end{enumerate}

\subsection*{Attention Check}
\label{att}

These attention checks were shown below the corresponding definitions to ensure participants read how the terms ``trustworthy AI'' and ``reliable AI'', respectively, were meant to be understood (for the definitions, see \nameref{defs}).

\subsubsection*{Trust Group}

\textbf{Before continuing, please select only the statements below that align with the key components of trust in AI.}

\begin{enumerate}
    \item If the AI is trustworthy, it will do the right thing for the right ethical reasons.
    \item If the AI is trustworthy, it neither conceptualizes moral principles nor does it care about ethical norms.
    \item Trustworthy AI cares about ethical norms and/or conceptualizes moral principles.
    \item Trustworthy AI has nothing to do with the AI having moral principles or ethical norms.
\end{enumerate}

\subsubsection*{Reliable Group}

\textbf{Before continuing, please select only the statements below that align with the key components of reliability in AI.}

\begin{enumerate}
    \item If the AI is reliable, it will produce the expected outcome for the tasks it was assigned.
    \item If the AI is reliable, it will neither produce the expected outcome, nor is it likely to do so in the future.
    \item Reliable AI consistently produces the expected outcome.
    \item Reliable AI has nothing to do with the AI being effective or consistent.
\end{enumerate}

\subsection*{Vignettes}
\label{vign}

Participants were instructed to read three distinct vignettes related to automotive AI: Planning Assistance, Parking Assistance, and Steering Assistance. Each vignette was designed to evaluate AI accountability, AI blameworthiness, confidence in using the AI, and confidence in learning to drive. The vignettes were identical except for the key terms ``trustworthy'' and ``reliable,'' as specified by the participants' group assignment. The key terms were not in bold as they are below, but italicized.

\subsubsection*{Planning Assistance}
\label{vign_plan}

Picture yourself in the driver's seat of a modern vehicle. You're stuck in the midst of heavy traffic during your daily commute. The vehicle is equipped with a \textbf{trustworthy/reliable} AI route-planning system, designed by the manufacturer to analyze real-time traffic data and recommend a new route to avoid congestion. You \textbf{trust/rely} on the suggestion, and the AI switches lanes and diverts to the suggested route without any additional input from you, the driver.

\subsubsection*{Parking Assistance}
\label{vign_park}

Visualize yourself navigating a suburban neighborhood in your vehicle equipped with \textbf{trustworthy/reliable} AI parking assistance. You come across a tight parallel parking space, and the AI recommends activating the \textbf{trustworthy/reliable} AI parking assistance. You \textbf{trust/rely} on the AI to park the vehicle, and it maneuvers the car into the tight parking spot without any additional input from you, the driver.

\subsubsection*{Steering Assistance}
\label{vign_steer}

Imagine driving on a rural road with your car's \textbf{trustworthy/reliable} AI driving assistance providing real-time safety alerts. As you approach a curve, the \textbf{trustworthy/reliable} AI uses its sensors to detect an obstacle ahead and displays a warning on your dashboard screen. It also suggests that the AI should steer the vehicle to bypass the obstacle. You \textbf{trust/rely} on the suggestion, and the AI steers the vehicle to avoid the obstacle without any additional input from you, the driver.

\subsection*{Experience}
\label{exp}

Participants were asked to check any of the following boxes that correspond to their experience with autonomous automotive technology. For the purpose of this study, 'autonomous' refers to the vehicle's ability to perform specific actions (such as planning a route, maintaining lane position, parking the vehicle, or steering) independently of the driver. Checking any of these boxes indicated that the participant was assigned to the 'Experience' category.

\begin{itemize}
    \item I have experience driving a vehicle with autonomous technology for lane-keeping assistance.
    \item I have experience driving a vehicle with autonomous technology for route-planning assistance.
    \item I have experience driving a vehicle with autonomous technology for parking assistance.
    \item I have experience driving a vehicle with autonomous technology for steering assistance.
    \item I have had an automobile accident while driving a vehicle equipped with one of the autonomous technologies listed above.
\end{itemize}

\subsection*{Technology Acceptance Model Questionnaire}
\label{TAM_Qs}

The items in the Technology Acceptance Model questionnaire were presented in a random order to prevent biasing effects. Both groups, ``Trustworthy AI'' and ``Reliable AI,'' answered the questionnaire. The TAM questionnaire is based on Choung et al\cite{bib14}. Titles of the constructs were not shown to participants.

\begin{enumerate}
    \item \textbf{Perceived Ease of Use}: Learning to use automotive AI has been or would be easy for me.
    \item \textbf{Perceived Usefulness}: Using automotive AI has improved or would improve my performance at accomplishing driving-related tasks.
    \item \textbf{Intention to Use}: Using automotive AI is something I would do in the future or will continue to do so.
    \item \textbf{Ability Trust}: Automotive AI technologies are competent in their area of expertise.
    \item \textbf{Benevolence Trust}: Automotive AI technologies care about our well-being.
    \item \textbf{Integrity Trust}: Automotive AI technologies do not abuse the information and advantage they have over their users.
    \item \textbf{General Trust}: I trust that automotive AI can offer information and services that are in my best interest.
    \item \textbf{Attitude}: I feel positive toward automotive AI technologies.
\end{enumerate}

\subsection*{Supplementary Analysis}

Additional analyses were conducted to explore the impact of demographic variables, such as gender, age, and experience with AI, on the main outcomes. These analyses were considered exploratory, as specified in the pre-registration document (Section 8.1), which states that any analyses not part of the primary hypotheses but providing additional insights would be exploratory. Specifically, we examined interactions between demographic variables (gender, age, and experience) and the dependent variables. An ordinal regression model was employed to assess the influence of demographic factors. The model was fit using the \texttt{clmm} function, which allows for the inclusion of both fixed and random effects. The model specification was as follows: \texttt{ordinal\_model <- clmm(response \textasciitilde{} demo\_variable + (1 \textbar{} participant), data = current\_data)}.

\subsection*{Gender}
\label{gender}

In this analysis, we examined the impact of gender on various attitudes towards automotive AI technologies. Given the sparsity of data in certain gender categories, ``Non-Binary,'' ``Other,'' and ``Prefer not to say'' were combined into a single ``Other'' category to ensure more stable estimates. We utilized ordinal regression models to assess the relationship between gender and respondents' perceptions across the various questions asked above, the eight TAM questions and the four vignette questions. The models included 478 observations, and we report the coefficients and confidence intervals for the predictor variables. Below are the detailed results for each question, including the specific model fit statistics, coefficients, and confidence intervals for the predictor coefficients.

\subsubsection*{AI Accountability}

For AI Accountability, being female was associated with a significant decrease in AI accountability ratings (Estimate = \(-0.836\), SE = \(0.253\), \(z\) = \(-3.31\), \(p\) = \(0.000930\), CI = \([-1.33, -0.341]\)). The ``Other'' category did not show a significant effect (Estimate = \(-1.60\), SE = \(1.66\), \(z\) = \(-0.962\), \(p\) = \(0.336\), CI = \([-4.85, 1.66]\)).

\subsubsection*{AI Blameworthiness}

For AI Blameworthiness, females reported lower blameworthiness ratings (Estimate = \(-0.734\), SE = \(0.234\), \(z\) = \(-3.14\), \(p\) = \(0.0017\), CI = \([-1.19, -0.275]\)). The ``Other'' category did not show a significant effect (Estimate = \(1.53\), SE = \(1.61\), \(z\) = \(0.951\), \(p\) = \(0.341\), CI = \([-1.62, 4.68]\)).

\subsubsection*{Confidence in Driving}

For Confidence in Driving, there was a significant decrease in confidence in driving for females (Estimate = \(-0.785\), SE = \(0.290\), \(z\) = \(-2.71\), \(p\) = \(0.0067\), CI = \([-1.35, -0.217]\)). The ``Other'' category showed a marginally significant effect (Estimate = \(-2.40\), SE = \(1.41\), \(z\) = \(-1.71\), \(p\) = \(0.088\), CI = \([-5.16, 0.357]\)).

\subsubsection*{Confidence in Learning}

For Confidence in Learning, confidence in learning was significantly lower for females (Estimate = \(-0.932\), SE = \(0.242\), \(z\) = \(-3.85\), \(p\) = \(0.000117\), CI = \([-1.41, -0.458]\)). The ``Other'' category also showed a significant negative effect (Estimate = \(-3.41\), SE = \(1.54\), \(z\) = \(-2.21\), \(p\) = \(0.027\), CI = \([-6.44, -0.383]\)). \newline

In summary, the analyses indicate that being female significantly negatively affects perceptions in several areas, including AI accountability, AI blameworthiness, confidence in driving, and confidence in learning. The combined ``Other'' category also showed significant negative effects in some areas, though often with larger confidence intervals, reflecting greater variability due to the lower number of observations in these groups.

\subsubsection*{Technology Acceptance Questions}

These results do not include Question 3 ``Using automotive AI is something I would do in the future or will continue to do so'' as this is discussed above under Intention to Use.

\subsubsection*{Question 1: Learning to use automotive AI would be easy for me}

The model had a log-likelihood of \(-619.43\) and an AIC of \(1250.86\). The coefficient for the \texttt{Female} variable was estimated at \(-0.8482\) with a standard error of \(0.1749\), resulting in a \(z\)-value of \(-4.849\) and a \(p\)-value of \(1.24 \times 10^{-6}\), indicating a statistically significant negative effect of being female on the perception of ease of learning. The coefficient for the \texttt{Other} variable was estimated at \(-1.6893\) with a standard error of \(0.7398\), resulting in a \(z\)-value of \(-2.283\) and a \(p\)-value of \(0.0224\), indicating a statistically significant negative effect for the combined ``Other'' category. The 95\% confidence intervals for these estimates ranged from \(-1.19\) to \(-0.507\) for \texttt{Female} and from \(-3.17\) to \(-0.207\) for \texttt{Other}.

Additionally, for Question 1, the interaction effect between being female and having no experience with automotive AI was found to be significant, revealing a significant decrease in the ease of learning ratings (Estimate = \(-0.8611\), SE = \(0.3469\), \(z\) = \(-2.482\), \(p\) = \(0.0131\), CI = \([-1.54, -0.183]\)). The ``No Experience'' category was defined by including participants who reported no experience with any of the listed AI automotive technologies (see \nameref{exp} for the questions). Specifically, this category comprises 257 participants, while 221 participants reported having experience with these technologies. This suggests that females with no experience find it significantly harder to learn to use automotive AI compared to their male counterparts with no experience. This model included only 471 observations, excluding data points from the ``Other'' category due to insufficient sample size, with a log-likelihood of \(-597.38\) and an AIC of \(1208.77\). The model converged after six iterations, with a maximum gradient of \(1.61 \times 10^{-13}\) and a condition number of \(57\).

\subsubsection*{Question 2: Using automotive AI would improve my performance at accomplishing driving-related tasks}

The model had a log-likelihood of \(-687.65\) and an AIC of \(1387.30\). The coefficient for the \texttt{Female} variable was estimated at \(-0.2253\) with a standard error of \(0.1672\), resulting in a \(z\)-value of \(-1.348\) and a \(p\)-value of \(0.178\), indicating that the effect of gender on performance improvement is not statistically significant. The coefficient for the \texttt{Other} variable was estimated at \(-0.1820\) with a standard error of \(0.7251\), resulting in a \(z\)-value of \(-0.251\) and a \(p\)-value of \(0.802\), also indicating no significant effect. The 95\% confidence intervals for these estimates ranged from \(-0.554\) to \(0.102\) for \texttt{Female} and from \(-1.62\) to \(1.28\) for \texttt{Other}.

\subsubsection*{Question 3: Using automotive AI is something I would do in the future or will continue to do so.}

The model had a log-likelihood of \(-699.62\) and an AIC of \(1411.24\). The coefficient for the \texttt{Female} variable was estimated at \(-0.5015\) with a standard error of \(0.1678\), resulting in a \(z\)-value of \(-2.988\) and a \(p\)-value of \(0.0028\), indicating a statistically significant negative effect of being female on the intention to use automotive AI. The coefficient for the \texttt{Other} variable was estimated at \(-1.3177\) with a standard error of \(0.7724\), resulting in a \(z\)-value of \(-1.706\) and a \(p\)-value of \(0.0880\), indicating a marginally significant negative effect for the combined ``Other'' category. The 95\% confidence intervals for these estimates ranged from \(-0.832\) to \(-0.174\) for \texttt{Female} and from \(-2.82\) to \(0.290\) for \texttt{Other}.

\subsubsection*{Question 4: Automotive AI technologies are competent in their area of expertise}

The model had a log-likelihood of \(-621.03\) and an AIC of \(1254.05\). The coefficient for the \texttt{Female} variable was estimated at \(-0.2867\) with a standard error of \(0.1717\), resulting in a \(z\)-value of \(-1.670\) and a \(p\)-value of \(0.0949\), indicating a marginally significant negative effect of being female on the perception of AI competence. The coefficient for the \texttt{Other} variable was estimated at \(-0.6836\) with a standard error of \(0.7779\), resulting in a \(z\)-value of \(-0.879\) and a \(p\)-value of \(0.3795\), indicating no significant effect for the combined ``Other'' category. The 95\% confidence intervals for these estimates ranged from \(-0.624\) to \(0.0492\) for \texttt{Female} and from \(-2.20\) to \(0.927\) for \texttt{Other}.

\subsubsection*{Question 5: Automotive AI technologies care about our well-being}

The model had a log-likelihood of \(-716.91\) and an AIC of \(1445.81\). The coefficient for the \texttt{Female} variable was estimated at \(-0.2455\) with a standard error of \(0.1654\), resulting in a \(z\)-value of \(-1.484\) and a \(p\)-value of \(0.138\), indicating that the effect of gender on the perception of information abuse is not statistically significant. The coefficient for the \texttt{Other} variable was estimated at \(-1.0647\) with a standard error of \(0.7535\), resulting in a \(z\)-value of \(-1.413\) and a \(p\)-value of \(0.158\), also indicating no significant effect for the combined ``Other'' category. The 95\% confidence intervals for these estimates ranged from \(-0.570\) to \(0.0783\) for \texttt{Female} and from \(-2.58\) to \(0.457\) for \texttt{Other}.

\subsubsection*{Question 6: Automotive AI technologies do not abuse the information and advantage they have over their users}

The model had a log-likelihood of \(-666.04\) and an AIC of \(1344.08\). The coefficient for the \texttt{Female} variable was estimated at \(-0.1793\) with a standard error of \(0.1686\), resulting in a \(z\)-value of \(-1.064\) and a \(p\)-value of \(0.288\), indicating that the effect of gender on trust in AI is not statistically significant. The coefficient for the \texttt{Other} variable was estimated at \(-0.6749\) with a standard error of \(0.7119\), resulting in a \(z\)-value of \(-0.948\) and a \(p\)-value of \(0.343\), also indicating no significant effect for the combined ``Other'' category. The 95\% confidence intervals for these estimates ranged from \(-0.510\) to \(0.151\) for \texttt{Female} and from \(-2.10\) to \(0.745\) for \texttt{Other}.

\subsubsection*{Question 7: I trust that automotive AI can offer information and service that is in my best interest}

The model had a log-likelihood of \(-646.52\) and an AIC of \(1305.03\). The coefficient for the \texttt{Female} variable was estimated at \(-0.2345\) with a standard error of \(0.1712\), resulting in a \(z\)-value of \(-1.369\) and a \(p\)-value of \(0.1709\), indicating that the effect of gender on positive feelings toward AI is not statistically significant. The coefficient for the \texttt{Other} variable was estimated at \(-1.4584\) with a standard error of \(0.7831\), resulting in a \(z\)-value of \(-1.862\) and a \(p\)-value of \(0.0625\), indicating a marginally significant negative effect for the combined ``Other'' category. The 95\% confidence intervals for these estimates ranged from \(-0.571\) to \(0.101\) for \texttt{Female} and from \(-2.98\) to \(0.182\) for \texttt{Other}.

\subsubsection*{Question 8: I feel positive toward automotive AI technologies}

The model had a log-likelihood of \(-683.79\) and an AIC of \(1379.58\). The coefficient for the \texttt{Female} variable was estimated at \(-0.5831\) with a standard error of \(0.1695\), resulting in a \(z\)-value of \(-3.439\) and a \(p\)-value of \(0.000583\), indicating a statistically significant negative effect of being female on the perception of AI trustworthiness. The coefficient for the \texttt{Other} variable was estimated at \(-1.6211\) with a standard error of \(0.7767\), resulting in a \(z\)-value of \(-2.087\) and a \(p\)-value of \(0.036871\), indicating a statistically significant negative effect for the combined ``Other'' category. The 95\% confidence intervals for these estimates ranged from \(-0.917\) to \(-0.252\) for \texttt{Female} and from \(-3.13\) to \(-0.00372\) for \texttt{Other}.

\subsection*{Age}
\label{age}

We conducted ordinal regression analyses to examine the relationship between age groups and responses to various questions about AI. The analysis included three age groups: 68 participants aged 18-30, 240 participants aged 30-45, 166 participants aged 45-65, and 3 participants who preferred not to disclose their age. The results for each main question are summarized below.

\subsubsection*{AI Accountability}
The analysis for the ``AI Accountability'' question encountered issues with convergence. Specifically, the model fitting process produced NaN values for standard errors, statistics, and confidence intervals. This issue indicates instability in parameter estimation, likely due to a highly significant deviation from a uniform distribution of responses. A chi-square goodness-of-fit test confirmed that the response distribution for this question significantly deviated from a uniform distribution (\(X^2 = 438.24\), \(\text{df} = 4\), \(p < 2.2 \times 10^{-16}\)). This non-uniform distribution contributed to the model's inability to accurately estimate parameters for the age groups. Consequently, no reliable estimates or confidence intervals could be reported for the ``AI Accountability'' question.

\subsubsection*{AI Blameworthiness}
For the ``AI Blameworthiness'' question, the age group 30-45 years old had a significant negative association with the response (estimate = \(-0.510\), SE = \(0.237\), \(z = -2.15\), \(p = 0.031\), \(95\%\ CI = [-0.974, -0.0456]\)), indicating that this age group was less likely to attribute blameworthiness to AI compared to the reference group (18-30 years old). Similarly, the age group 45-65 years old also showed a significant negative association (estimate = \(-0.665\), SE = \(0.277\), \(z = -2.40\), \(p = 0.016\), \(95\%\ CI = [-1.21, -0.123]\)).

\subsubsection*{Confidence in Driving}
For the ``Confidence in Driving'' question, neither the age group 30-45 years old (estimate = \(0.587\), SE = \(0.429\), \(z = 1.37\), \(p = 0.171\), \(95\%\ CI = [-0.254, 1.43]\)) nor the age group 45-65 years old (estimate = \(0.00446\), SE = \(0.449\), \(z = 0.00993\), \(p = 0.992\), \(95\%\ CI = [-0.875, 0.884]\)) showed significant associations with the response.

\subsubsection*{Confidence in Learning}
For the ``Confidence in Learning'' question, the age group 30-45 years old exhibited a significant positive association with the response (estimate = \(1.34\), SE = \(0.474\), \(z = 2.83\), \(p = 0.00469\), \(95\%\ CI = [0.411, 2.27]\)), indicating higher confidence in AI learning capabilities. The age group 45-65 years old did not show a significant association (estimate = \(0.184\), SE = \(0.493\), \(z = 0.373\), \(p = 0.709\), \(95\%\ CI = [-0.782, 1.15]\)).\newline

The analysis highlights distinct age-related differences in perceptions of AI accountability, blameworthiness, and confidence in AI applications. For ``AI Blameworthiness,'' both the 30-45 (estimate = \(-0.510\), \(p = 0.031\)) and 45-65 (estimate = \(-0.665\), \(p = 0.016\)) age groups were significantly less likely to attribute blame to AI compared to the 18-30 age group. In contrast, age did not significantly impact ``Confidence in Driving.'' However, for ``Confidence in Learning,'' the 30-45 age group exhibited a significant positive association (estimate = \(1.34\), \(p = 0.00469\)), indicating higher confidence in AI learning capabilities. These findings underscore that age significantly influences perceptions of AI blameworthiness and learning confidence, while its impact on driving confidence is not significant.

\subsubsection*{Technology Acceptance Questions}
These results do not include Question 3 ``Using automotive AI is something I would do in the future or will continue to do so.'', as this is discussed above under Intention to Use.

\subsubsection*{Question 1: Learning to use automotive AI would be easy for me}
The model included 478 observations, with a log-likelihood of \(-628.39\) and an AIC of \(1270.77\). The model converged after six iterations, with a maximum gradient of \(3.91 \times 10^{-13}\) and a condition number of \(192\). For the age group 30-45 years old, the estimate was \(0.5453\) (SE = \(0.2569\), \(z = 2.122\), \(p = 0.0338\), \(95\%\ CI = [0.0413, 1.05]\)), indicating a significant effect. For the age group 45-65 years old, the estimate was \(0.2459\) (SE = \(0.2681\), \(z = 0.917\), \(p = 0.3590\), \(95\%\ CI = [-0.280, 0.772]\)). For those who preferred not to say their age, the estimate was \(-1.3716\) (SE = \(0.9424\), \(z = -1.455\), \(p = 0.1456\), \(95\%\ CI = [-3.31, 0.519]\)).

\subsubsection*{Question 2: Using automotive AI would improve my performance at accomplishing driving-related tasks}
The model included 478 observations, with a log-likelihood of \(-686.95\) and an AIC of \(1387.89\). The model converged after five iterations, with a maximum gradient of \(2.25 \times 10^{-8}\) and a condition number of \(210\). For the age group 30-45 years old, the estimate was \(-0.1411\) (SE = \(0.2541\), \(z = -0.555\), \(p = 0.5788\), \(95\%\ CI = [-0.641, 0.357]\)). For the age group 45-65 years old, the estimate was \(-0.1599\) (SE = \(0.2688\), \(z = -0.595\), \(p = 0.5520\), \(95\%\ CI = [-0.688, 0.367]\)). For those who preferred not to say their age, the estimate was \(-1.6808\) (SE = \(0.9138\), \(z = -1.839\), \(p = 0.0659\), \(95\%\ CI = [-3.53, 0.180]\)).

\subsubsection*{Question 3: Using automotive AI is something I would do in the future or will continue to do so.}
The model included 478 observations, with a log-likelihood of \(-699.54\) and an AIC of \(1413.09\). The model converged after five iterations, with a maximum gradient of \(1.24 \times 10^{-9}\) and a condition number of \(230\). For the age group 30-45 years old, the estimate was \(0.3453\) (SE = \(0.2492\), \(z = 1.386\), \(p = 0.1659\)). For the age group 45-65 years old, the estimate was \(-0.1084\) (SE = \(0.2584\), \(z = -0.420\), \(p = 0.6748\)). For those who preferred not to say their age, the estimate was \(-1.7513\) (SE = \(0.8848\), \(z = -1.979\), \(p = 0.0478\)). The results indicate that age did not significantly impact the intention to use automotive AI for most age groups, except for those who preferred not to disclose their age (three participants), who showed a significant negative association.

\subsubsection*{Question 4: Automotive AI technologies are competent in their area of expertise}
The model included 478 observations, with a log-likelihood of \(-620.11\) and an AIC of \(1254.21\). The model converged after six iterations, with a maximum gradient of \(3.42 \times 10^{-13}\) and a condition number of \(230\). For the age group 30-45 years old, the estimate was \(0.2917\) (SE = \(0.2598\), \(z = 1.123\), \(p = 0.2614\), \(95\%\ CI = [-0.219, 0.801]\)). For the age group 45-65 years old, the estimate was \(0.2389\) (SE = \(0.2742\), \(z = 0.871\), \(p = 0.3847\), \(95\%\ CI = [-0.297, 0.774]\)). For those who preferred not to say their age, the estimate was \(-1.7949\) (SE = \(1.0631\), \(z = -1.688\), \(p = 0.0913\), \(95\%\ CI = [-4.06, 0.275]\)).

\subsubsection*{Question 5: Automotive AI technologies care about our well-being}
The model included 478 observations, with a log-likelihood of \(-716.66\) and an AIC of \(1447.31\). The model converged after five iterations, with a maximum gradient of \(7.36 \times 10^{-9}\) and a condition number of \(290\). For the age group 30-45 years old, the estimate was \(-0.2750\) (SE = \(0.2503\), \(z = -1.099\), \(p = 0.2720\), \(95\%\ CI = [-0.768, 0.215]\)). For the age group 45-65 years old, the estimate was \(-0.2872\) (SE = \(0.2629\), \(z = -1.093\), \(p = 0.2746\), \(95\%\ CI = [-0.804, 0.227]\)). For those who preferred not to say their age, the estimate was \(-1.8193\) (SE = \(0.9600\), \(z = -1.895\), \(p = 0.0581\), \(95\%\ CI = [-3.92, 0.0660]\)).

\subsubsection*{Question 6: Automotive AI technologies do not abuse the information and advantage they have over their users}
The model included 478 observations, with a log-likelihood of \(-663.44\) and an AIC of \(1340.87\). The model converged after five iterations, with a maximum gradient of \(1.72 \times 10^{-7}\) and a condition number of \(190\). For the age group 30-45 years old, the estimate was \(0.4084\) (SE = \(0.2619\), \(z = 1.559\), \(p = 0.1189\), \(95\%\ CI = [-0.105, 0.923]\)). For the age group 45-65 years old, the estimate was \(0.6435\) (SE = \(0.2741\), \(z = 2.348\), \(p = 0.0189\), \(95\%\ CI = [0.107, 1.18]\)), indicating a significant effect. For those who preferred not to say their age, the estimate was \(-0.6537\) (SE = \(0.9484\), \(z = -0.689\), \(p = 0.4906\), \(95\%\ CI = [-2.55, 1.28]\)).

\subsubsection*{Question 7: I trust that automotive AI can offer information and service that is in my best interest}
The model included 478 observations, with a log-likelihood of \(-646.03\) and an AIC of \(1306.05\). The model converged after six iterations, with a maximum gradient of \(1.08 \times 10^{-13}\) and a condition number of \(260\). For the age group 30-45 years old, the estimate was \(0.1024\) (SE = \(0.2622\), \(z = 0.390\), \(p = 0.6962\), \(95\%\ CI = [-0.414, 0.615]\)). For the age group 45-65 years old, the estimate was \(0.0647\) (SE = \(0.2725\), \(z = 0.237\), \(p = 0.8123\), \(95\%\ CI = [-0.471, 0.598]\)). For those who preferred not to say their age, the estimate was \(-2.2544\) (SE = \(1.0182\), \(z = -2.214\), \(p = 0.0268\), \(95\%\ CI = [-4.44, -0.257]\)), indicating a significant effect.

\subsubsection*{Question 8: I feel positive toward automotive AI technologies}
The model included 478 observations, with a log-likelihood of \(-687.67\) and an AIC of \(1389.34\). The model converged after five iterations, with a maximum gradient of \(1.19 \times 10^{-7}\) and a condition number of \(300\). For the age group 30-45 years old, the estimate was \(0.0266\) (SE = \(0.2546\), \(z = 0.104\), \(p = 0.9170\), \(95\%\ CI = [-0.474, 0.525]\)). For the age group 45-65 years old, the estimate was \(-0.1137\) (SE = \(0.2666\), \(z = -0.426\), \(p = 0.6699\), \(95\%\ CI = [-0.638, 0.408]\)). For those who preferred not to say their age, the estimate was \(-2.4890\) (SE = \(0.9929\), \(z = -2.507\), \(p = 0.0122\), \(95\%\ CI = [-4.63, -0.539]\)), indicating a significant effect.\newline

In summation, significant age-related effects were observed in several questions. Participants aged 30-45 found it significantly easier to learn automotive AI (\(p = 0.0338\)). The 45-65 age group showed significantly higher trust that automotive AI would not abuse its advantages (\(p = 0.0189\)). These findings suggest that younger and middle-aged adults are more accepting of automotive AI. Additionally, participants who preferred not to say their age showed significantly lower trust that automotive AI can offer information and service in their best interest (\(p = 0.0268\)) and felt significantly less positive toward automotive AI technologies (\(p = 0.0122\)).

\subsection*{Experience}

We conducted ordinal regression analyses to examine the relationship between the experience of respondents and their responses to different questions regarding AI. The ``No Experience'' category was defined by including participants who reported no experience with any of the listed AI automotive technologies (see \nameref{exp} for the exact questions). Specifically, this category comprises 257 participants, while 221 participants reported having experience with these technologies. The analysis compared individuals with no experience versus those with experience with automotive AI. The results are summarized below for each main question.

\subsubsection*{AI Accountability}
For the ``AI Accountability'' question, respondents with no experience with AI showed a significant positive association with the response (estimate = \(0.918\), SE = \(0.338\), \(z = 2.71\), \(p = 0.007\), \(95\%\ \text{CI} = [0.255, 1.58]\)). This indicates that individuals without AI experience were more likely to have a higher level of accountability assigned to AI compared to those with AI experience.

\subsubsection*{AI Blameworthiness}
Similarly, for the ``AI Blameworthiness'' question, respondents with no experience with AI also showed a significant positive association with the response (estimate = \(0.865\), SE = \(0.313\), \(z = 2.77\), \(p = 0.006\), \(95\%\ \text{CI} = [0.252, 1.48]\)). This suggests that those without AI experience were more likely to attribute blameworthiness to AI compared to those with AI experience.

\subsubsection*{Confidence in Driving}
In contrast, for the ``Confidence in Driving'' question, respondents with no experience with AI exhibited a significant negative association with the response (estimate = \(-0.740\), SE = \(0.287\), \(z = -2.58\), \(p = 0.010\), \(95\%\ \text{CI} = [-1.30, -0.177]\)). This indicates that individuals without AI experience had lower confidence in AI driving capabilities compared to those with AI experience.

\subsubsection*{Confidence in Learning}
For the ``Confidence in Learning'' question, respondents with no experience with AI again showed a significant negative association with the response (estimate = \(-1.00\), SE = \(0.320\), \(z = -3.13\), \(p = 0.002\), \(95\%\ \text{CI} = [-1.63, -0.376]\)). This suggests that those without AI automotive experience had lower confidence in AI learning capabilities compared to those with experience.\newline

The analysis revealed distinct patterns in attitudes toward AI based on experience. Respondents with no experience with AI assigned significantly higher levels of accountability and blameworthiness to AI, with estimates of \(0.918\) (SE = \(0.338\), \(z = 2.71\), \(p = 0.007\)) and \(0.865\) (SE = \(0.313\), \(z = 2.77\), \(p = 0.006\)) respectively, indicating a greater tendency to hold AI accountable and blameworthy. Conversely, these individuals displayed significantly lower confidence in AI's driving and learning capabilities, reflected in negative estimates of \(-0.740\) (SE = \(0.287\), \(z = -2.58\), \(p = 0.010\)) and \(-1.00\) (SE = \(0.320\), \(z = -3.13\), \(p = 0.002\)), suggesting that lack of AI experience correlates with skepticism regarding AI's effectiveness in these domains. These findings underscore the impact of AI experience on perceptions of accountability, blameworthiness, and confidence in AI capabilities.

\subsection*{Technology Acceptance Questions}
These results do not include Question 3 ``Using automotive AI is something I would do in the future or will continue to do so.'', as this is discussed above under Intention to Use.

\subsubsection*{Question 1: Learning to use automotive AI would be easy for me}
The model included \(478\) observations, with a log-likelihood of \(-622.10\) and an AIC of \(1254.21\). The estimate for participants with no experience was \(-0.7957\) (SE = \(0.1736\), \(z = -4.583\), \(p < 0.001\), \(95\%\ \text{CI} = [-1.14, -0.457]\)), indicating a significant negative effect.

\subsubsection*{Question 2: Using automotive AI would improve my performance at accomplishing driving-related tasks}
The model included \(478\) observations, with a log-likelihood of \(-681.40\) and an AIC of \(1372.79\). The estimate for participants with no experience was \(-0.6345\) (SE = \(0.1686\), \(z = -3.762\), \(p < 0.001\), \(95\%\ \text{CI} = [-0.967, -0.305]\)), indicating a significant negative effect.

\subsubsection*{Question 3: Using automotive AI is something I would do in the future or will continue to do so.}
The model included \(478\) observations, with a log-likelihood of \(-691.40\) and an AIC of \(1392.79\). The estimate for participants with no experience was \(-0.8802\) (SE = \(0.1709\), \(z = -5.151\), \(p < 0.001\), \(95\%\ \text{CI} = [-1.22, -0.547]\)), indicating a significant negative effect for participants with no experience using automotive AI.

\subsubsection*{Question 4: Automotive AI technologies are competent in their area of expertise}
The model included \(478\) observations, with a log-likelihood of \(-617.41\) and an AIC of \(1244.81\). The estimate for participants with no experience was \(-0.557\) (SE = \(0.173\), \(z = -3.22\), \(p = 0.001\), \(95\%\ \text{CI} = [-0.898, -0.219]\)), indicating a significant negative effect.

\subsubsection*{Question 5: Automotive AI technologies care about our well-being}
The model included \(478\) observations, with a log-likelihood of \(-713.86\) and an AIC of \(1437.72\). The estimate for participants with no experience was \(-0.5179\) (SE = \(0.1659\), \(z = -3.121\), \(p = 0.002\), \(95\%\ \text{CI} = [-0.844, -0.194]\)), indicating a significant negative effect.

\subsubsection*{Question 6: Automotive AI technologies do not abuse the information and advantage they have over their users}
The model included \(478\) observations, with a log-likelihood of \(-665.56\) and an AIC of \(1341.12\). The estimate for participants with no experience was \(-0.279\) (SE = \(0.168\), \(z = -1.66\), \(p = 0.097\), \(95\%\ \text{CI} = [-0.609, 0.0499]\)), indicating a non-significant effect.

\subsubsection*{Question 7: I trust that automotive AI can offer information and service that is in my best interest}
The model included \(478\) observations, with a log-likelihood of \(-647.23\) and an AIC of \(1304.45\). The estimate for participants with no experience was \(-0.3007\) (SE = \(0.1707\), \(z = -1.762\), \(p = 0.078\), \(95\%\ \text{CI} = [-0.636, 0.0332]\)), indicating a non-significant effect.

\subsubsection*{Question 8: I feel positive toward automotive AI technologies}
The model included \(478\) observations, with a log-likelihood of \(-685.25\) and an AIC of \(1380.50\). The estimate for participants with no experience was \(-0.5721\) (SE = \(0.1688\), \(z = -3.39\), \(p < 0.001\), \(95\%\ \text{CI} = [-0.904, -0.243]\)), indicating a significant negative effect.

\subsection*{Discussion of Exploratory Analysis}
The exploratory analysis highlighted several significant demographic effects on attitudes towards automotive AI, focusing on gender, age, and experience.

\subsubsection*{Gender}
Gender was a notable factor influencing attitudes towards automotive AI. Female participants generally reported lower levels of AI accountability amd AI blameworthiness, suggesting female participants are more lenient toward AI than their male counterparts. Additionally, female participants generally reported lower confidence in driving with AI and confidence in learning to drive with AI compared to their male counterparts. Specifically, being female had a significant negative effect on the intention to use automotive AI, with a coefficient of \(-0.5015\) (SE = \(0.1678\), \(z = -2.988\), \(p = 0.0028\)). What is more, an interaction effect between being female and having no AI experience was significant for the ease of learning to use automotive AI (Estimate = \(-0.8611\), SE = \(0.3469\), \(z = -2.482\), \(p = 0.0131\)), indicating that females with no AI experience found it significantly more challenging compared to their male counterparts with no experience.

\subsubsection*{Age}
Age significantly influenced perceptions of AI in various ways. Older participants were less likely to attribute blame or accountability to automotive AI. Specifically, the 30-45 and 45-65 age groups showed a lower likelihood of attributing blame to AI (estimates of \(-0.510\), \(p = 0.031\) and \(-0.665\), \(p = 0.016\), respectively) compared to the 18-30 age group. Conversely, the 30-45 age group reported higher confidence in AI learning capabilities (estimate = \(1.34\), \(p = 0.00469\)). However, age did not significantly affect confidence in driving.

\subsubsection*{Experience}
Experience with AI was a strong determinant of attitudes towards automotive AI. Participants with no experience assigned significantly higher levels of accountability and blameworthiness to AI, with estimates of \(0.918\) (SE = \(0.338\), \(z = 2.71\), \(p = 0.007\)) and \(0.865\) (SE = \(0.313\), \(z = 2.77\), \(p = 0.006\)), respectively. They also exhibited lower confidence in driving and learning to drive with AI, reflected in estimates of \(-0.740\) (SE = \(0.287\), \(z = -2.58\), \(p = 0.010\)) and \(-1.00\) (SE = \(0.320\), \(z = -3.13\), \(p = 0.002\)). These findings highlight that a lack of AI experience correlates with more negative perceptions, an increased likelihood to blame, and lower acceptance of AI technologies across various aspects, including ease of learning, perceived performance improvement, and overall positivity towards AI.

Finally, the analysis revealed no significant interaction effects between any demographic variables and the group label ('trustworthy AI'). This indicates that the relationship between the demographic factors and the respondents' perceptions was consistent across both the 'trustworthy AI' and 'reliable AI' groups.

\subsubsection*{Conclusion}

The exploratory analysis highlights significant demographic effects on attitudes toward automotive AI, providing further insight into the philosophical debate regarding AI trust. Specifically, the finding that female participants and participants with no experience report lower levels of confidence stands in stark contrast with the absence of interaction effects between group assignment and gender or experience. This suggests that the label ``trustworthy AI'' does improve the negative attitudes held by these participants. Moreover, since these participants also report lower ratings for intention to use, this further emphasizes the viewpoint that trustworthy AI is an inappropriate target for alleviating algorithm aversion. Consequently, the exploratory analysis indicates that demographic factors, particularly gender and experience, significantly influence perceptions of AI's trustworthiness and reliability. The negative association between being female and various trust metrics suggests that AI design and labeling strategies should be sensitive to these demographic differences. This finding supports the ethical argument that AI systems should be designed and communicated in ways that recognize and address the varied trust dynamics among users.

For developers and policymakers, the exploratory analysis provides crucial insights into how demographic factors shape user attitudes toward AI. The significant negative impact of being female on trust-related metrics implies that marketing and educational strategies should be tailored to address the specific concerns of female users. By focusing on reliability and transparency, developers can better manage user expectations and develop more inclusive acceptance of automotive AI technologies.

\end{document}